\documentclass[onecolumn,prd,aps,amsmath,showpacs,amssymb]{revtex4}
\advance\oddsidemargin by -\textwidth
\advance\oddsidemargin by -1in
\evensidemargin=\oddsidemargin
\def\fsl#1{\setbox0=\hbox{$#1$}           % set a box for #1 
   \dimen0=\wd0                                 % and get its size
   \setbox1=\hbox{/} \dimen1=\wd1               % get size of /
   \ifdim\dimen0>\dimen1                        % #1 is bigger
      \rlap{\hbox to \dimen0{\hfil/\hfil}}      % so center / in box
      #1                                        % and print #1
   \else                                        % / is bigger
      \rlap{\hbox to \dimen1{\hfil$#1$\hfil}}   % so center #1
      /                                         % and print /
   \fi}    
\makeatletter
\def\@maketitle{\newpage
 \null
 {\normalsize \tt \begin{flushright} 
  \begin{tabular}[t]{l} \@date 
  \end{tabular}
 \end{flushright}}
 \begin{center}
 \vskip 2em
 {\LARGE \@title \par} \vskip 2.5em {\large \lineskip .5em
 \begin{tabular}[t]{c}\@author 
 \end{tabular}\par} 
 \end{center}
 \par
 \vskip 1.5em} 
\makeatother
\usepackage{graphicx}
\usepackage{color}
\usepackage{amsmath,amssymb}
\usepackage{array}

\newcommand{\vev}[1]{\langle #1 \rangle}
\newcommand{\lessim}{\hspace{0.3em}\raisebox{0.4ex}{$<$}\hspace{-0.75em}\raisebox{-.7ex}{$\sim$}\hspace{0.3em}}

\newcommand{\TeV}{\text{TeV}}
\newcommand{\GeV}{\text{GeV}}

\newcommand{\crit}{{\rm crit}}
\newcommand{\tr}{\mbox{tr}}
\newcommand{\gtwo}{I\kern-.1em I\,}
\newcommand{\be}{\begin{equation}}
\newcommand{\ee}{\end{equation}}
\newcommand{\beq}{\begin{eqnarray}}
\newcommand{\eeq}{\end{eqnarray}}
\newcommand{\bpm}{\begin{pmatrix}}
\newcommand{\epm}{\end{pmatrix}}
\newcommand{\cl}{\, \rm C.L.}
\newcommand{\blacktriangles}{\raisebox{0.1ex}{\rotatebox[origin=c]{0}{\small $\blacktriangle$}}\,}
\newcommand{\blackdiamond}{\raisebox{0.22ex}{\rotatebox[origin=c]{45}{\scriptsize $\blacksquare$}}\,}
\newcommand{\blackcircles}{\raisebox{-0.045ex}{\rotatebox[origin=c]{0}{\Large $\bullet$}}\,}
\newcommand{\whitecircles}{\!\raisebox{-0.045ex}{\rotatebox[origin=c]{0}{\Large $\circ$}}\,}
\newcommand{\drawsquare}[2]{\hbox{%
\rule{#2pt}{#1pt}\hskip-#2pt%  left vertical
\rule{#1pt}{#2pt}\hskip-#1pt%  lower horizontal
\rule[#1pt]{#1pt}{#2pt}}\rule[#1pt]{#2pt}{#2pt}\hskip-#2pt%  upper horizontal
\rule{#2pt}{#1pt}}% right vertical
\newcommand{\Ysymm}{\raisebox{-.5pt}{\drawsquare{6.5}{0.4}}\hskip-0.4pt%
        \raisebox{-.5pt}{\drawsquare{6.5}{0.4}}}%  symmetric second rank
\begin{document}

\title{A hybrid 4$^{\textrm{th}}$ generation: Technicolor with top-seesaw}

\author{Hidenori S. Fukano}
\email{hidenori.f.sakuma@jyu.fi} 
\author{Kimmo Tuominen}
\email{kimmo.i.tuominen@jyu.fi}
\affiliation{Department of Physics, University of Jyv\"askyl\"a, P.O.Box 35, FIN-40014 Jyv\"askyl\"a, Finland \\
and 
Helsinki Institute of Physics, P.O.Box 64, FIN-00014 University of Helsinki, Finland\\}

\begin{abstract}
We consider a model combining technicolor with the top quark condensation. As a concrete model
for Technicolor we use the Minimal Walking Technicolor, and this will result in the appearance of a novel fourth generation whose leptons constitute a usual weak doublet while the QCD quarks are vectorlike singlets under the weak interactions. We carry out an analysis of the mass spectra and precision measurement constraints, and find the model viable. We contrast the model with present LHC data and 
discuss the future prospects. 
\end{abstract}

\maketitle
%%%%%%%%%%%%%%%%%%%%%%%%%%%%%%%%%%%%%%%%%%
%%%%%%%%%%%%%%%%%%%%%%%%%%%%%%%%%%%%%%%%%%
%%%%%%%%%%%%%%%%%%%%%%%%%%%%%%%%%%%%%%%%%%
%
%
%
\section{Introduction}
%
%%%%%%%%%%%%%%%%%%%%%%%%%%%%%%%%%%%%%%%%%%
%%%%%%%%%%%%%%%%%%%%%%%%%%%%%%%%%%%%%%%%%%

The Standard Model (SM) of the elementary particle interactions is known to describe nature at the level of current observations. However, SM is believed to be an incomplete theory, mainly, because we can not explain the origin of the observed mass patterns of the matter fields, how many generations of the matter fields there are or why there is en excess of matter over antimatter. There are many continuing efforts to work out these mysteries. One possible paradigm is to explain the Higgs mechanism in the SM is by the strong coupling gauge theory  dynamics. 
Along this line, technicolor (TC) was proposed \cite{Weinberg:1975gm} (for reviews, see \cite{Hill:2002ap,Sannino:2009za,Sannino:2008ha}). Here the electroweak symmetry breaking is due to the condensation of new matter fields called technifermions. Unfortunately, the old fashioned but simple TC model  based on the QCD-like gauge theory dynamics is incompatible with the electroweak precision data at the LEP experiments \cite{Peskin:1990zt}. Recently, we have introduced TC models that are compatible with the electroweak precision data \cite{Sannino:2004qp,Dietrich:2005jn}. In these TC models, technifermions transform in higher representation under the TC gauge group, and these models naturally lead to a walking behavior \cite{Holdom:1984sk,Yamawaki:1985zg,AkibaYanagida,AppelquistKarabaliWijewardhana}
 of the technicolor coupling constant, thanks to a nontrivial infrared fixed point \cite{Banks:1981nn}.  Many groups are studying via first principle calculations whether the nonperturbative gauge theory dynamics underlying the proposed walking TC models  \cite{Catterall:2007yx,Hietanen:2008mr,DelDebbio:2008zf,Hietanen:2009az,Bursa:2009we,Shamir:2008pb,DeGrand:2008kx,DeGrand:2010na,Fodor:2009ar,Kogut:2010cz}. Among several such walking TC models one of the most promising candidates is the minimal walking Technicolor (MWT) model. The matter content of MWT is two Dirac flavors of technifermions  in the adjoint representation of the SU$(2)$ TC gauge group. When gauged under the electroweak interactions, one must cancel the Witten anomaly arising due to the odd number of techniquark doublets. In order to cancel this anomaly, a fourth generation of leptons is introduced into the model. With this particle content, the MWT model overcomes the obstacles arising from the electroweak precision tests \cite{Dietrich:2005jn}. 
%Moreover, if we assign the $U(1)_Y$ charges for the technifermions in the MWT model as the same charge of the SM quarks, we need to add the fourth generations leptons in order to cancel the Witten anomaly. 
%So it is instructive to study a model based on the MWT in order to reveal mysteries of the nature which are the origin of the mass and the number of 
It is interesting to note that perhaps anomaly cancellation in a more complete underlying theory may provide hints on the number of fermion generations \cite{Antipin:2010it}.

For explaining various mass patterns of the known matter fields within a TC framework, a well-known approach is the extended TC (ETC) \cite{Dimopoulos:1979es}, in which the technifemions and the SM fermions are embedded into a larger gauge group (${\cal G}_{\rm ETC}$). In ETC, after ${\cal G}_{\rm ETC}$ breaks down to the TC gauge group and the technifermion condensation is triggered by the TC gauge dynamics, the SM fermions obtain their masses via the massive ETC gauge bosons coupled with the technifermion condensates. If an ETC gauge group breaks sequentially, this model may explain the observed mass hierarchies of the SM fermions \cite{Appelquist:1993sg,Appelquist:2003hn}. However, it is hard to explain a large top quark mass, or more precisely, a top-bottom mass splitting. To address this particular issue, an alternative to ETC, the top quark condensation model, was proposed in a form of the low energy effective model  \cite{Miransky:1988xi,Nambu89,Marciano89,Bardeen:1989ds}. This model was then completed by a topcolor model where, generically, a gauge group $SU(3)_{\rm QCD} \times U(1)_Y$ of the SM is typically extended to ${\cal G}_{\rm topC} = $SU(3)$_1 \times $SU(3)$_2 \times $U(1)$_{Y1} \times $U(1)$_{Y2}$ \cite{Hill:1991at}. A model which combines TC/ETC and topcolor dynamics has been proposed \cite{Hill:1994hp}, and several groups are pursuing model building along this line \cite{Fukano:2008iv,Ryttov:2010fu}. We can regard this type of models as a kind of low scale TC \cite{Eichten:1979ah,Lane:1989ej,Lane:2009ct,Delgado:2010bb}, which may explain \cite{Eichten:2011sh} the observed excess in the dijet invariant mass in the $W +$ jets events reported by the CDF collaboration \cite{Aaltonen:2011mk}.

Recently we have proposed a new possibility of a model, in which we combined the MWT model and the top quark condensation \cite{Fukano:2011fp}. This model is one of top seesaw assisted TC models. The dynamics of top quark condensation was taken to be similar as in the top seesaw model \cite{Dobrescu:1997nm,Chivukula:1998wd,He:2001fz}. A main new feature in our model with respect to earlier work is that in addition to explaining the mass patterns of the heavy SM fermions, it predicts a {\it hybrid} fourth generation. This means that the new leptons transform as chiral fermions under the electroweak gauge group, but there are also new QCD quarks transforming as vector-like fermions under this group. 

Many aspects of a sequential fourth generation have been studied earlier \cite{Frampton:1999xi}.  For example, it has been shown that a sequential fourth generation can be accommodated by the electroweak precision data \cite{He:2001tp,Kribs:2007nz,Chanowitz:2009mz,Erler:2010sk,Eberhardt:2010bm,Chanowitz:2010bm} and that the fourth generation quarks help to describe the current experimental data on CP violation and rare decays of B mesons better within the CKM-paradigm \cite{Bobrowski:2009ng,Soni:2010xh,Buras:2010pi,Buras:2010nd,Buras:2010cp}. The fourth generation leptons have been discussed in various contexts  \cite{Buras:2010cp,Carpenter:2010sm,Burdman:2009ih,Blennow:2010th,Antusch:2008tz,Antusch:2006vwa,Frandsen:2009fs,Lenz:2011gd,Masina:2011ew}. Finally, it has been pointed out that the sequential fourth generation model can trigger the electroweak symmetry breaking \cite{Holdom:2006mr,Holdom:2009rf,Hashimoto:2009ty,Hashimoto:2010at}. However, a non-sequential fourth generation is a less investigated option, which would arise naturally due to internal consistency of some other beyond the SM sector \cite{Antipin:2010it,Knochel:2011ng}.
In the previous work \cite{Fukano:2011fp}, we have described the dynamics of this hybrid fourth generation model and discussed several constraints coming from the experiments. In this paper we will formulate the dynamics in the language of the effective theory which will allow us to compare the hybrid fourth generation model with the current electroweak precision data and discuss also recent results from the LHC.

%%%%%%%%%%%%%%%%%%%%%%%%%%%%%%%%%%%%%%%%%%
%%%%%%%%%%%%%%%%%%%%%%%%%%%%%%%%%%%%%%%%%%
%%%%%%%%%%%%%%%%%%%%%%%%%%%%%%%%%%%%%%%%%%
%
%
%
\section{An overview of the model and dynamics}
%
%%%%%%%%%%%%%%%%%%%%%%%%%%%%%%%%%%%%%%%%%%
%%%%%%%%%%%%%%%%%%%%%%%%%%%%%%%%%%%%%%%%%%

We consider a model termed model B in \cite{Fukano:2011fp}, which combines the MWT model and top quark condensation. The SU(3)$_c\times$U(1)$_Y$ of SM is in this model embedded into ${\cal G}=$SU(3)$_1\times$SU(3)$_2\times$U(1)$_{Y1}\times$U(1)$_{Y2}$. To see how the fourth generation matter fields originate, consider the charge assignments of the technicolor fields:
%In our model, we consider a model based on the minimal walking technicolor 
In MWT the SU(2)$_L$ doublet technifermion $({\cal Q}_L)$ has $Y({\cal Q}_L) = 1/6$ under the U(1)$_{Y2}$ gauge symmetry. Therefore, in order to cancel the Witten and gauge anomalies in the technicolor sector, we add one SM-like SU(2)$_L$ doublet of leptons. Moreover, in this model, the third generation quarks are assumed to obtain their masses only by the seesaw mechanism after some condensations are triggered, but other quarks obtain their masses mainly by ETC interactions with the technifermion condensates. This mechanism is the same as in the top quark seesaw model \cite{Dobrescu:1997nm,Chivukula:1998wd,He:2001fz}. When we concentrate on the non-technicolored new matter sector, there are one SM-like {\it  chiral} lepton doublet and two {\it vector-like} extra quarks. In this sense, we call it a {\it hybrid fourth generation} -model. A most minimal model consists of particles in Table \ref{particle-model}. Techniquarks of the MWT sector are denoted by ${\cal Q}_L$, ${\cal T}_R$ and ${\cal B}_R$, while the fourth generation leptons, required to exist by the internal consistency of the MWT model, by $L^{(4)}_L$, $N^{(4)}_R$ and $E^{(4)}_R$.   
The usual SM fields are $Q^{(3)}/L^{(3)}$ denoting the third generation and $Q^{(i)}/L^{(i)}\,\,(i =1,2)$ the first and second generation quarks/leptons; the ``SM'' in the charge assignments shown in the table represents the ordinary SM charge values.  Finally, the fields $U^{(4)}$ and $D^{(4)}$ are the fourth generation QCD quarks transforming as vector-like fermions under the EW gauge symmetry. 

%Now, we extend the ordinary SU(3)$_{\rm QCD} \times $U(1)$_Y$ gauge group to ${\cal G} = $SU(3)$_1 \times $SU(3)$_2 \times $U(1)$_{Y1} \times $U(1)$_{Y2}$ which we assume to break according to the pattern ${\cal G} \to $SU(3)$_{\rm QCD} \times $U(1)$_Y$. 
We consider the case that SU(3)$_1$ gauge coupling $(h_1)$ is stronger than SU(3)$_2$ gauge coupling $(h_2)$, i.e. the ratio $\cot \theta = h_1/h_2$ is larger than one. Furthermore, we consider the U(1)$_{Y1}$ gauge coupling $(h'_1)$ to be stronger than the U(1)$_{Y2}$ gauge coupling $(h'_2)$,  which implies that the ratio $\cot \theta' = h'_1/h'_2 $ is larger than one. In the notation introduced above, the QCD coupling and hypercharge U(1)$_Y$ couplings are given by  $g_{\rm QCD} = h_1 \sin \theta = h_2 \cos \theta$ and $g_Y = h'_1 \sin \theta' = h'_2 \cos \theta'$.

\begin{table}[h]
\begin{center}
\begin{tabular}{| c || c | c | c | c | c | c |}
\hline
field & SU(2)$_{\rm TC} $ & SU(3)$_1$  & SU(3)$_2$ & SU(2)$_L$ & U(1)$_{Y1}$ 
& U(1)$_{Y2}$ 
\\
\hline 
${\cal Q}_L$ & $\Ysymm$ & 1 & 1 & 2 & 0 & 1/6 
\\
${\cal T}_R$ & $\Ysymm$ & 1 & 1 & 1 & 0 & 2/3 
\\
${\cal B}_R$ & $\Ysymm$ & 1 & 1 & 1 & 0 & -1/3 
\\ 
\hline
\, & \, & \, & \, & \, & \, & \, 
\\[-2.5ex] 
\hline
$L^{(4)}_L$ & 1 & 1 & 1 & 2 & 0 & -1/2
\\
$N^{(4)}_R$ & 1 & 1 & 1 & 1 & 0 & 0
\\ 
$E^{(4)}_R$ & 1 & 1 & 1 & 1 & 0 & -1 
\\ 
\hline
\, & \, & \, & \, & \, & \, & \, 
\\[-2.5ex] 
\hline
$Q^{(3)}_L$ & 1 & 3 & 1 & 2 & 1/6 & 0  
\\
$U^{(3)}_R$ & 1 & 1 & 3 & 1 & 0 & 2/3 
\\
$D^{(3)}_R$ & 1 & 1 & 3 & 1 & 0 & -1/3 
\\
\hline 
\, & \, & \, & \, & \, & \, & \, 
\\[-2.5ex] 
\hline 
$L^{(3)}_L$ & 1 & 1 & 1& 2 & -1/2 & 0 
\\
$N^{(3)}_R$ & 1 & 1 & 1 & 1 & 0 & 0
\\ 
$E^{(3)}_R$ & 1 & 1 & 1& 1 & -1 & 0 
\\ 
\hline 
\, & \, & \, & \, & \, & \, & \, 
\\[-2.5ex]
\hline
$U^{(4)}_L$ & 1 & 1 & 3 & 1 & 0 & 2/3  
\\
$U^{(4)}_R$ & 1 & 3 & 1 & 1 & 2/3 & 0 
\\
$D^{(4)}_L$ & 1 & 1 & 3 & 1 & 0 & -1/3 
\\
$D^{(4)}_R$ & 1 & 3 & 1 & 1 & -1/3 & 0 
\\
\hline 
\, & \, & \, & \, & \, & \, & \, 
\\[-2.5ex] 
\hline 
$Q^{(1,2)}$ & 1 & 1 & 3 & SM & 0 & SM
\\
$L^{(1,2)}$ & 1 & 1 &1 & SM & 0 & SM
\\
\hline
\end{tabular}
\caption{Particle content and charge assignments of the model. $Q/L^{(1,2,3)}$ are the first,second and third generations quarks/leptons of the SM, and ``SM'' in the columns for quantum numbers stands for the ordinary SM charge. All fermions are represented in terms of the weak gauge eigenbasis.} 
\label{particle-model}
\end{center}
\end{table}%

The breaking pattern ${\cal G} \to $SU(3)$_{\rm QCD} \times $U(1)$_Y$, assumed to occur at some energy $\mu\gg v_{\textrm{weak}}$, leads to the appearance of 8 + 1 massive gauge bosons. The eight massive gauge bosons associated with the breaking of SU(3)$_1\times$ SU(3)$_2$ are called ``colorons '' and denoted with $G'$. The one massive gauge boson associated with the breaking of U(1)$_1\times$U(1)$_2$ is denoted with $Z'$. Their masses are denoted, respectively by $M_{G^\prime}$ and $M_{Z^\prime}$. At low energies, the interactions via colorons or $Z'$ exchange lead to effective four fermion interactions which we will write down explicitly below. Then we can divide the model into three parts, which are the SM part, the MWT part and the four fermion interaction part. We concentrate on the four fermion sector, and we neglect the first and second family fermions for simplicity. We assume that  the masses of all leptons in this model as well as the masses of light quarks are explained by some underlying ETC dynamics operating at much higher scales and we aim to explain dynamically only the mass patterns of the third and fourth quark generations. 

The exchange of a heavy coloron, $G'$, gives 
\beq
{\cal L}^{4f}_{G'} =
-\frac{4\pi \kappa_3}{2M^2_{G'}} \!\!
\sum^{SU(3)_1}_{f,f'} \!\!\!
\left( \bar{f} \gamma^\mu  T^a f \right)\left( \bar{f'} \gamma_\mu T^a f' \right)
-\frac{4\pi }{2M^2_{G'}} \frac{\alpha^2_{\rm QCD}}{\kappa_3} \!\!
\sum^{SU(3)_2}_{f,f'} \!\!\!
\left( \bar{f} \gamma^\mu  T^a f \right)\left( \bar{f'} \gamma_\mu T^a f' \right)
\,,
\label{4f-3}
\eeq
where $f \in SU(3)_i$ stands for $SU(3)_i$ charged fermions, 
while the exchange of a heavy $Z'$ gives 
\beq
{\cal L}^{4f}_{Z'} =
-\frac{4\pi \kappa_1}{2M^2_{Z'}} 
\sum_{f,f'} \left( Y^{(f)}_1 \bar{f} \gamma^\mu f \right)\left( Y^{(f')}_1 \bar{f'} \gamma_\mu f' \right)
-\frac{4\pi}{2M^2_{Z'}} \frac{\alpha^2_Y}{\kappa_1}
\sum_{f,f'} \left( Y^{(f)}_2 \bar{f} \gamma^\mu f \right)\left( Y^{(f')}_2 \bar{f'} \gamma_\mu f' \right)
\,, 
\label{4f-1}
\eeq
where $Y^{(f)}_i$ shows hypercharge of $f$ under the $U(1)_{Yi}$. 
Now we concentrate on the parts derived from the more strongly interacting sectors $SU(3)_1$ and $U(1)_{Y1}$, and we assume that the contributions from the more weakly interacting gauge sector is negligible in comparison to the stronger sector contributions.

Applying Fiertz rearrangements to Eqs. (\ref{4f-3}) and (\ref{4f-1}), we obtain the four fermion interactions which can be divided into two parts as 

\beq
{\cal L}^{4f} = {\cal L}^{4f}_S + {\cal L}^{4f}_V\,,
\eeq
where 
\beq
{\cal L}^{4f}_S 
= 
\left[ G_{3S} + \frac{1}{9} G_{1S} \right] \left(\overline{U^{(4)}_R}\, Q^{(3)}_L \right)^2
+
\left[ G_{3S} - \frac{1}{18} G_{1S} \right] \left(\overline{D^{(4)}_R}\, Q^{(3)}_L\right)^2
+
\frac{1}{2} G_{1S}  \left(\overline{E^{(3)}_R}\, L^{(3)}_L\right)^2
\,,
\label{4f-scalar}
\eeq
and 
\beq
{\cal L}^{4f}_V
=
&& \hspace*{-3ex}
-\left[ G_{3V} + \frac{1}{36} G_{1V} \right] 
\left\{
\left(\overline{Q^{(3)}_L}\, \gamma_\mu\, Q^{(3)}_L \right)^2
+ 2\left(\overline{U^{(3)}_L}\, \gamma^\mu\, D^{(3)}_L \right)\left(\overline{D^{(3)}_L}\, \gamma_\mu\, U^{(3)}_L \right)
\right\}
\nonumber\\[1ex]
&& \hspace*{-3ex}
-\left[ G_{3V} + \frac{4}{9} G_{1V} \right] \left(\overline{U^{(4)}_R}\, \gamma_\mu\, U^{(4)}_R \right)^2 
-\left[ G_{3V} + \frac{1}{9} G_{1V} \right] \left(\overline{D^{(4)}_R}\, \gamma_\mu\, D^{(4)}_R \right)^2 
\nonumber\\[1ex]
&& \hspace*{-3ex}
-2 \left[ G_{3V} - \frac{2}{9} G_{1V} \right] \left(\overline{U^{(4)}_R}\, \gamma^\mu\, D^{(4)}_R \right) \left(\overline{D^{(4)}_R}\, \gamma_\mu\, U^{(4)}_R \right)\,.
\label{4f-vector}
\eeq
Here the four-fermion couplings are given in terms of the parameters of the underlying theory as
\beq
G_{3S} = 4 G_{3V}= \frac{8}{9}\frac{4\pi \kappa_3}{M^2_{G'}} %\frac{3^2-1}{3^2}
\quad,\quad
G_{1S} =4 G_{1V} = \frac{8\pi \kappa_1}{M^2_{Z'}}\,.
\label{GS-GV-relations}
\eeq

To analyse the formation of a dynamical condensate in this model we concentrate on the interaction in Eq. (\ref{4f-scalar}); the interaction in Eq. (\ref{4f-vector}) will contribute to the analysis of electroweak precision constraints. Regarding the interactions among third family leptons, it is enough to consider the scalar four fermion interactions in Eq. (\ref{4f-scalar}) since we will treat them as massless in the calculation of electroweak precision constraints.  

The analysis of the condensate formation is based on the conventional Nambu-Jona-Lasinio (NJL) model treatment. Generally, a scalar four fermion interaction implies the gap equation
\beq
\Sigma =  \frac{2 N G_S \Lambda^2}{(4\pi)^2} \Sigma \left[ 1 - \frac{\Sigma^2}{\Lambda^2} \ln \frac{\Lambda^2+\Sigma^2}{\Sigma^2}\right]\,,
\label{gapeq-NJL}
\eeq
where $\Sigma$ is the dynamical mass of fermions and $N = 3$ or $1$corresponding to the interaction derived from the SU(3) or U(1) gauge boson exchange, respectively. The ultraviolet scale below which the four fermions interactions are applicable is denoted by $\Lambda$, and we set $\Lambda = M_{G'} = M_{Z'}$ in this paper. To see how the critical coupling arises, we rewrite Eq. (\ref{gapeq-NJL}) as 
\beq
\frac{\Sigma^3}{\Lambda^2} \ln \frac{\Lambda^2+\Sigma^2}{\Sigma^2}
=
\Sigma \left( \frac{1}{g_\crit} - \frac{1}{g}\right)\,,
\label{gapeq-NJL-mod}
\eeq
where $g \equiv 2 N \Lambda^2 G/(4\pi)^2 $ and $g_\crit = 1$. 
One can easily see that the l.h.s in Eq. (\ref{gapeq-NJL-mod}) is always positive, while the r.h.s can be positive only if $g>g_\crit$.
Therefore, if $g<g_\crit$, the gap equation Eq.(\ref{gapeq-NJL}) does not have any dynamical mass solutions; it only has the trivial solution $\Sigma =0$. On the other hand,  if $g >g_\crit$ the gap equation has a non-trivial dynamical solution $\Sigma \neq 0$. 

In the present model, from Eq. (\ref{4f-scalar}), the gap equations are given by
\beq
\Sigma_U=
g^{(34)}_U 
\Sigma_U \left[ 1 - \frac{\Sigma^2_U}{\Lambda^2} \ln \frac{\Lambda^2+\Sigma^2_U}{\Sigma^2_U}\right]
\quad , \quad
\Sigma_D= 
g^{(34)}_D  
\Sigma_D \left[ 1 - \frac{\Sigma^2_D}{\Lambda^2} \ln \frac{\Lambda^2+\Sigma^2_D}{\Sigma^2_D}\right]
\,,
\label{gapeq-present}
\eeq
where the dimensionless couplings are given by 
\beq
g^{(34)}_U = \frac{3}{2 \pi} \left[ \frac{8}{9}\kappa_3 + \frac{2}{27} \kappa_1 \right]
\quad,\quad
g^{(34)}_D= \frac{3}{2 \pi} \left[ \frac{8}{9}\kappa_3 - \frac{1}{27} \kappa_1 \right].
\label{dless-4f}
\eeq

The essential point, which leads to the desired seesaw mechanism for the heavy quark masses is, that in this model, after ${\cal G} \to $SU(3)$_{\rm QCD} \times $U(1)$_Y$ symmetry breaking, we are allowed to add SM gauge invariant mass terms 
\beq
{\cal L}^0_{\rm mass} 
=
-M^{(43)}_U \overline{U^{(4)}_L} U^{(3)}_R - M^{(44)}_U \overline{U^{(4)}_L}U^{(4)}_R 
-M^{(43)}_D \overline{D^{(4)}_L} D^{(3)}_R - M^{(44)}_D \overline{D^{(4)}_L}D^{(4)}_R 
+ {\rm h.c.}\,
\label{baremass-model2}
\eeq
Qualitatively, in order to realize the top-bottom mass splitting after the condensate formation, the parameters in Eq.(\ref{baremass-model2}) need to have some hierarchy; this is where the seesaw mechanism enters. In our model we find, after fitting the physical masses of top and bottom quarks that $M_U^{(44)}/m_t\sim \Lambda/(1 {\textrm{\,TeV}})$, $M_U^{(43)}/m_t\sim \Lambda/(1{\textrm{\,TeV}})$,
$M_D^{(44)}/m_b\sim 50 \Lambda/(1{\textrm{\,TeV}})$ and $M_D^{(44)}/m_b\sim \Lambda/(1{\textrm{\,TeV}})$, where $\Lambda$ is the UV cutoff of the theory where the effective four fermion interactions should be matched onto the massive gauge boson exchange in the underlying theory.

Since our approach is bottom-up model building, the origin of these terms is left unspecified. For example, the $M^{(44)}_{U/D}$ terms could arise from the vacuum expectation value of a scalar field in $(\bar{3},3)$ representation of SU(3)$_1\times$ SU(3)$_2$, while the $M^{(43)}_{U/D}$ terms could be induced by some underlying ETC interactions operating at yet higher scales in comparison to $M_{G^\prime}$ and $M_{Z^\prime}$ relevant for the breaking of SU(3)$_1\times$SU(3)$_2$.  

We also require that there is no condensation of third generation leptons. This channel is controlled by the coupling
\beq
 g_E^{(33)}= \frac{\Lambda^2}{8\pi^2}G_{1S}=\kappa_1.
\eeq

In order to set a theoretical constraint for $\kappa_3,\kappa_1$, we consider the criticality conditions of the couplings $g^{(34)}_U$, $g^{(34)}_D$  and $g_E^{(33)}$ together with the position of the Landau pole of the $U(1)_{Y_1}$ coupling, which we assume to be more strongly coupled of the two $U(1)$ factors. For this purpose we consider the renormalization group equation (RGE) for the $U(1)_{Y_1}$ gauge coupling. 
At one loop, the RGE is given by 
\be
\frac{d \alpha_{Y1}}{d \ln \mu} = \frac{b_{Y1}}{2\pi} \alpha^2_{Y1},
\ee
 with $b_{Y1} = 40/9$. So the running of $\alpha_{Y1}$ is 
\beq
\frac{1}{\alpha_{Y1}(\mu)} - \frac{1}{\alpha_{Y1}(\Lambda_{\rm UV})} = \frac{b_{Y1}}{2\pi} \ln \frac{\Lambda_{\rm UV}}{\mu}\,,
\eeq
where $\mu < \Lambda_{\rm UV}$. By definition, the Landau pole is reached at scale $\Lambda_L$, where  $1/\alpha_{Y1}(\Lambda_L) = 0$.  If we denote the scale of the symmetry breaking as $\Lambda  < \Lambda_L$, then 
\beq
\frac{1}{\alpha_{Y1}(\Lambda)} = \frac{b_{Y1}}{2\pi} \ln \frac{\Lambda_L}{\Lambda}\, .
\label{const-Landau-pole}
\eeq
Since the low energy four-fermion coupling $\kappa_1$ is related to the gauge couplings of the $U(1)_{Y_1}$ and $U(1)_{Y}$ groups as $\kappa_1 = \alpha_{Y1} - \alpha_Y $, Eq. (\ref{const-Landau-pole}) allows one to determine $\kappa_1$ for given $\Lambda_L/\Lambda$. 

\begin{figure}[htbp]
\begin{center}
\includegraphics{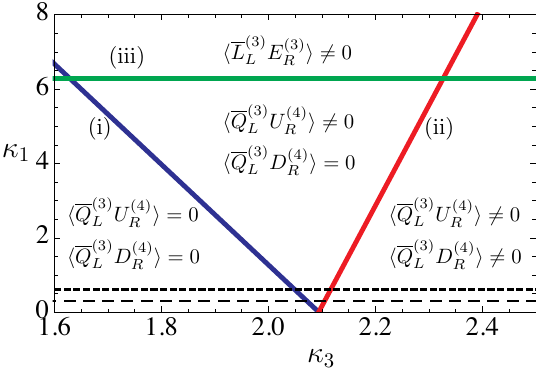} 
\caption{ 
The gap triangle for the present model.
The region above (i) represents $\vev{\overline{Q}_L^{(3)}U_R^{(4)}} \neq 0$, the region below (ii) represents $\vev{\overline{Q}_L^{(3)}D_R^{(4)}} \neq 0$ and the region above (iii) represents $\vev{\overline{L}_L^{(3)}E_R^{(3)}} \neq 0$. The condensates of $Q^{(3)}_L-U^{(4)}_R$ and $Q^{(3)}_L-D^{(4)}_R$ form their condensates in an area which is to the right of (ii) and below (iii). The two dashed lines represent constraints from a position of the Landau pole, Eq. (\ref{const-Landau-pole}), for $\Lambda_L/\Lambda=10,100$ corresponding to the upper and lower lines.
\label{gaptriangle-model}}
\end{center}
\end{figure}

The constraints on the parameter space $(\kappa_3, \kappa_1)$ are shown in Fig.\ref{gaptriangle-model}. To obtain the desired pattern of condensations, we need the criticality conditions $g^{(34)}_{U/D}> 1$ to hold. We want to avoid condensation of third generation leptons, and hence we require that $g_E^{(33)}<1$.  These conditions result in the gap-triangle shown in the figure. The dashed horizontal lines are determined by the constraint on the position of the Landau pole, Eq. (\ref{const-Landau-pole}).  

%%%%%%%%%%%%%%%%%%%%%%%%%%%%%%%%%%%%%%%%%%
%%%%%%%%%%%%%%%%%%%%%%%%%%%%%%%%%%%%%%%%%%
%%%%%%%%%%%%%%%%%%%%%%%%%%%%%%%%%%%%%%%%%%
%
\section{Effective Lagrangian and Mass Spectrum}
%
%%%%%%%%%%%%%%%%%%%%%%%%%%%%%%%%%%%%%%%%%%
%%%%%%%%%%%%%%%%%%%%%%%%%%%%%%%%%%%%%%%%%%

In this section we will apply the analysis of \cite{He:2001fz} to our model. We consider mixing of all generations, and assume the seesaw mechanism for the generation of fourth and third generation quark masses.
% and treat the seesaw mechanism under an assumption. 

%%%%%%%%%%%%%%%%%%%%%%%%%%%%%%%%%%%%%%%%%%
%
\subsection{Gap equation in general}
%
%%%%%%%%%%%%%%%%%%%%%%%%%%%%%%%%%%%%%%%%%%

In order to consider the dynamical contributions from all quark generations, 
we diagonalize the quark mass matrix as follows:
\beq
U^{(\alpha)}_{L/R} = U^{L/R}_{\alpha\beta} u^{(\beta)}_{L/R}\,,
\quad, \quad
D^{(\alpha)}_{L/R} = D^{L/R}_{\alpha\beta} d^{(\beta)}_{L/R}\,,
\label{quark-mixing}
\eeq
where $U^{L/R}_{\alpha\beta}$ and $D^{L/R}_{\alpha\beta}$ are unitary matices.
Here $u,d$ are the fermion fields in the mass eigenbasis and the corresponding mass eigenvalues will be written as $m_{ui}$ and $m_{di}$. In terms of the mass basis, the gap equations, Eq. (\ref{gapeq-present}), are 
\beq
\Sigma_U
&=& 
g^{(34)}_U
\sum^4_{\alpha=1} {\rm Re}\left[ U^L_{3\alpha} U^{R*}_{4\alpha}\right] 
m_{u\alpha} \left[ 1 - \frac{m^2_{u\alpha}}{\Lambda^2} \ln\frac{\Lambda^2+m^2_{u\alpha} }{m^2_{u\alpha}}\right]
\label{gapeq-U-af}
\,\\
\Sigma_D
&=& 
g^{(34)}_D
\sum^4_{\alpha=1} {\rm Re}\left[ D^L_{3\alpha} D^{R*}_{4\alpha}\right] 
m_{d\alpha} \left[ 1 - \frac{m^2_{d\alpha}}{\Lambda^2} \ln\frac{\Lambda^2+m^2_{d\alpha}}{m^2_{d\alpha}}\right]\,.
\label{gapeq-D-af}
\eeq

%%%%%%%%%%%%%%%%%%%%%%%%%%%%%%%%%%%%%%%%%%
%
\subsection{Low energy effective Lagrangian}
\label{LEEFT-higgs}
%
%%%%%%%%%%%%%%%%%%%%%%%%%%%%%%%%%%%%%%%%%%

Since the dynamical aspects of the top seesaw sector are treated in the NJL model, we can use the fermion bubble sum approximation with ${\cal L}^{4f}_S$ in Eq.(\ref{4f-scalar})~\cite{Bardeen:1989ds,He:2001fz} to obtain a low energy effective Lagrangian valid for $\mu < \Lambda \simeq M_{G'} \simeq M_{Z'}$. In this section, we consider the low energy effective Lagrangian for the top seesaw sector in accordance with \cite{He:2001fz} and its mixing with the Technicolor sector.

First, we introduce the auxiliary higgs fields $\Phi^{(0)}_{1,2}$ as  $\Phi^{(0)}_1 \sim \bar{D}^{(4)}_RQ^{(3)}_L$ and $\Phi^{(0)}_2 \sim \bar{U}^{(4)}_RQ^{(3)}_L$. Then, by using these auxiliary fields we can rewrite ${\cal L}^{4f}_S$ as 
\beq
{\cal L}^{\rm higgs}_{\mu=\Lambda}
=
- \Lambda^2 \left[ |\Phi^{(0)}_1|^2 + |\Phi^{(0)}_2|^2 \right] 
- \left[ y_{10} \bar{Q}^{(3)}_L \Phi^{(0)}_1 D^{(4)}_R + y_{20} \bar{Q}^{(3)}_L \tilde{\Phi}^{(0)}_2 U^{(4)}_R 
+ {\rm h.c. }\right]\,,
\eeq
where $\tilde{\Phi}_i \equiv i \tau^2 \Phi^*_i$, $y_{10} = [G_{3S} + (1/9)G_{1S}] \Lambda^2$ and $y_{20} = [G_{3S} - (1/18)G_{1S}]\Lambda^2$. Now we parametrize $\Phi^{(0)}_i\,(i=1,2)$ as
\beq
\Phi^{(0)}_i = \bpm \pi^+_{i0} \\[1ex] \dfrac{1}{\sqrt{2}} \left[ h^0_{i0} - i \pi^0_{i0} \right]\epm\,.
\label{higgs-paramet}
\eeq
So, in this picture, if the criticality condition $g^{(34)}_{U/D} >1$ is satisfied, the fields $\Phi^{(0)}_i$ have non-zero vacuum expectation values, $\vev{\Phi^{(0)}_i} = \left( 0 \,, \vev{h^0_{i0}}/\sqrt{2} \right)^T \neq 0$. This is obviously equivalent to $\Sigma_{U/D} \neq 0$. Now one can easily see ${\cal L}^{4f}_S + {\cal L}^0_{\rm mass}$ i.e. ${\cal L}^{\rm higgs}_{\mu=\Lambda} + {\cal L}^0_{\rm mass}$ has the Peccei-Quinn global $U(1)_A$ symmetry, under which each fermion transforms as \cite{He:2001fz}
\beq
&&
Q^{(3)}_L \to e^{-i\alpha}Q^{(3)}_L \,,\, U^{(3)}_R \to e^{i\alpha} U^{(3)}_R \,,\, D^{(3)}_R \to e^{i\alpha} D^{(3)}_R\,,\nonumber \\[1ex]
&&
U^{(4)}_{L/R} \to e^{i\alpha}U^{(4)}_{L/R} \,,\, D^{(4)}_{L/R} \to e^{i\alpha}D^{(4)}_{L/R}\,,\nonumber \\[1ex]
&&
\Phi^{(0)}_2 \to e^{-2i\alpha} \Phi^{(0)}_2\,,\,\Phi^{(0)}_1 \to e^{-2i\alpha} \Phi^{(0)}_1\,.
\eeq
If this $U(1)_A$ were an exact symmetry, its dynamical breaking by $\Sigma_{U/D} \neq 0$ would result in an appearance of a problematic massless Nambu-Goldstone boson. In the framework of topcolor model, this symmetry is explicitly broken by the topcolor instanton effect \cite{Hill:1994hp}. This can be represented by a four fermion interaction of the form
\beq
{\cal L}_A =
\frac{\xi h^2_1}{\Lambda^2} \epsilon^{ab} \left( \bar{U}^{(4)}_R Q^{(3)}_{L,a} \right) \left( \bar{D}
^{(4)}_R Q^{(3)}_{L,b} \right) + \rm h.c.\,,
\eeq
which can be rewritten using $\Phi^{(0)}_i$ as
\beq
{\cal L}_A
=
- \xi \Lambda^2 \left[\epsilon^{ab} \Phi^{(0)}_{1,a} \Phi^{(0)}_{2,b}\right] 
- \xi \sqrt{4\pi\kappa_3} 
\left\{
\left[ \epsilon^{ab} \bar{Q}^{(3)}_{L,a} \Phi^{(0)}_{1,b} \right] D^{(4)}_R 
+ \left[ \epsilon^{ab}\bar{Q}^{(3)}_{L,a} \tilde{\Phi}^{(0)}_{2,b} \right] U^{(4)}_R 
+ \rm h.c.
\right\}
\,
\eeq
Here $\epsilon^{ab}$ is the $SU(2)$ antisymmetric tensor and in section.\ref{numerical-results} we will set $\xi = 10^{-3}$ in accordance with \cite{He:2001fz}. Of course, this topcolor instanton effect affects also the gap equations. However if $\xi \leq 10^{-3}$, this effect is negligible \cite{He:2001fz}.
Therefore, the full Lagrangian at $\mu = \Lambda$ becomes ${\cal L}_\Lambda = {\cal L}^{\rm higgs}_{\mu = \Lambda} + {\cal L}_A$. 

Using the fermion bubble sum approximation \cite{Bardeen:1989ds,He:2001fz}, the low energy Lagrangian at 
$\mu < \lambda$ is
\beq
{\cal L}^{\rm higgs}
=
{\cal L}^{\rm higgs}_{\rm kin} (\Phi_1,\Phi_2)
- \left[ y_1 \bar{Q}^{(3)}_L \Phi_1 D^{(4)}_R + y_2 \bar{Q}^{(3)}_L \tilde{\Phi}_2 U^{(4)}_R 
+ {\rm h.c. }\right]
 - V(\Phi_1,\Phi_2)\,,
\label{reno-yukawa-H4G}
\eeq
where the potential $V(\Phi_1,\Phi_2)$ is given by \cite{He:2001fz}
\beq
V(\Phi_1,\Phi_2) 
&=& 
M^2_{11} |\Phi_1|^2 + M^2_{22} |\Phi_2|^2 
- M^2_{12} \left[ \Phi^\dagger_1 \Phi_2 + {\rm h.c.}\right] 
\nonumber\\[1ex]
&&
+ \lambda_b\left( \Phi^\dagger_1 \Phi_1 \right)^2 
+ \lambda_t \left( \Phi^\dagger_2 \Phi_2 \right)^2 
\nonumber\\[1ex]
&&
+ \lambda_{tb} \left[ \left( \Phi^\dagger_1 \Phi_1 \right) \left( \Phi^\dagger_2 \Phi_2 \right) - \left( \Phi^\dagger_1 \Phi_2 \right) \left( \Phi^\dagger_2 \Phi_1 \right)\right]
\nonumber\\[1ex]
&&
+ \left[ 
\lambda'_b \left( \Phi^\dagger_1 \Phi_1 \right) + \lambda'_t \left( \Phi^\dagger_2 \Phi_2 \right)
\right] 
\left[ \Phi^\dagger_1 \Phi_2 + {\rm h.c.} \right]\,.
\label{higgs-potential}
\eeq

Note that the higgs quartic coupling terms in Eq. (\ref{higgs-potential}) correspond to similar terms in a generic two Higgs doublets model \cite{Haber:1993an} after identifying the couplings in \cite{Haber:1993an} as $\lambda_1 \to \lambda_b,\lambda_2 \to \lambda_t,\lambda_3 = - \lambda_4 = \lambda_{tb},\lambda_5=0,\lambda_{6} \to \lambda'_b,\lambda_7 \to \lambda'_t$.
%

% \subsection{The PNGBs from the MWT sector}
Then consider the low energy effective theory of the pseudo Nambu-Goldstone bosons (PNGBs) from the MWT described by the chiral Lagrangian based on the $G/H = SU(4)/SO(4)$ \cite{Foadi:2007ue}. We decompose the non-linear sigma model field $U(x)$ for $G_{\rm global}/H_{\rm global}$ as
\beq
U_{\rm TC}(x) = \xi_{\rm TC}(x) \cdot E \cdot \xi^T_{\rm TC} (x) % = \xi^2_{\rm TC}(x) E
\quad,\quad
\xi(x)_{\rm TC} = \exp\left[\frac{i \Pi_{\rm TC}(x)}{f_{\rm TC}} \right]\,,
\eeq
where $\Pi_{\rm TC}(x) = \Pi^a(x) X^a\,,X^a \in {\cal G}-{\cal H} \,\,\,(a = 1, \cdots, 9)$ are NGB fields. 
Here  ${\cal G}$ and ${\cal H}$ denote the Lie algebra corresponding to $G$ and $H$, respectively and the decay constant for these NBGs is $f_{\rm TC} = v_{\rm TC}/\sqrt{2}$. 
%e used the relation $X^a E - E (X^a)^T = 0$ between broken generator $X^a$ and $E$. 
For the concrete expressions  of the generators $X^a$, see \cite{Foadi:2007ue}. 

To simplify the notation, we will not explicate the variables $x$ in what follows. The non-linear 
sigma fields $U_{\rm TC}, \xi_{\rm TC}$ transform as $U_{\rm TC} \to g  U_{\rm TC} g^T\,,\xi_{\rm TC} 
\to  g  \xi_{\rm TC}  g^T$ under $G$. The $4 \times 4$ matrix $E$ is given by
\beq
E = \bpm 0_{2 \times 2} & 1_{2 \times 2} \\ 1_{2 \times 2} & 0_{2 \times 2} \epm\,,
\eeq
which transforms $E \to g E g^T$ under $G$, and is invariant under $H$. The leading order chiral Lagrangian is
\beq
{\cal L}^{\rm PNGBs}_{\rm MWT}
=
\frac{f^2_{\rm TC}}{4} \tr \left| D_\mu U \right|^2\,,
\label{chilag}
\eeq
where the covariant derivative $D_\mu U_{\rm TC}$, is given by
\beq
D_\mu U_{\rm TC}
=
\partial_\mu U_{\rm TC} - i [g W^a_\mu L^a + g' Y B_\mu] U_{\rm TC} - i U_{\rm TC} [g W^a_\mu L^a + g' Y B_\mu]^T\,.
\eeq
Here $W^a_\mu,B_\mu$  are SM $SU(2)_L$ and $U(1)_Y$ gauge fields and $g,g'$ their gauge couplings. The $4 \times 4$ matrices $L^a$ and $Y$ are given by
\beq
L^a =  \bpm \sigma^a/2 & 0_{2 \times 2} \\ 0_{2 \times 2} & 0_{2 \times 2} \epm
\quad,\quad
Y=\bpm y/2 &\, &\, &\, \\ \,&y/2&\,&\,\\\,&\,&-(1+y)/2&\,\\\,&\,&\,&-(y-1)/2\epm\,,
\eeq
here $\sigma^a$ are the Pauli matrices, and $y/2$ is the hypercharge of the $SU(2)_L$ doublet technifermions .

Out of the nine NGBs ($\Pi_{\rm TC}$) in the MWT sector, three NGBs, denoted by 
$\pi_{\rm TC} = \pi^a_{\rm TC}X^a\,\,(a=1,2,3)$, will mix with the dynamical higgs bosons 
%$A^0,H^\pm$. 
%The latter 
which arise as the composite objects of the 3rd and 4th family quarks. In order to consider this mixing, 
%between $\pi_{\rm TC}$ and $A^0,H^\pm$, 
we first embed $\Phi_i$ in a form of $4 \times 4$ matrix as 
\beq
\Sigma_i=
\left( \hspace{-\arraycolsep}
\begin{array}{cccc}
0&0 & &\\
0&0 &\multicolumn{1}{l}{\raisebox{1.5ex}[0pt]{ {\parbox{12pt}{$\tilde{\Phi}_i$}}}}
&\multicolumn{1}{l}{\raisebox{1.2ex}[0pt]{ {\parbox{12pt}{$\Phi_i$}}}}
\\[1ex]
\multicolumn{2}{c}{\raisebox{0ex}[0pt]{ {\parbox{12pt}{${}^t\tilde{\Phi}_i$}}}}  &0&0\\[1ex]
\multicolumn{2}{c}{\raisebox{0ex}[0pt]{ {\parbox{12pt}{${}^t\Phi_i$}}}}& 0 &0  \\
\end{array}
\right)\,,
\eeq
where $\Phi_i \,(i=1,2)$ is as introduced in Eq.(\ref{higgs-potential}). We should note that $\Sigma_i$ 
transforms under not full $G = SU(4)$ but only $SU(2)_L \times SU(2)_R \subset G$.
Thus the mixing between the topseesaw sector and the MWT sector is \cite{Chivukula:2011ag}
\beq
V_M(U_{\rm TC},\Sigma_1,\Sigma_2)
\!\!\!&=&\!\!\!
%- \lambda_{\rm TC} \tr 
%\left( \Sigma^\dagger_{\rm TC} \Sigma_{\rm TC} - \frac{v^2_{\rm TC}}{2}\right)^2 
%\nonumber
%\\
%&&
- c_1 v^2_1 \tr \left| \Sigma_1- \frac{v_1}{\sqrt{2}}U_{\rm TC} \right|^2 
- c_2 v^2_2 \tr \left| \Sigma_2- \frac{v_2}{\sqrt{2}}U_{\rm TC} \right|^2\,,
\label{mixing-TSS-MWT}
\eeq
where $c_{1,2}\sim{\cal{O}}(1)$ are dimensionless parameters. 

The low energy effective theory is now specified by Eqs. (\ref{reno-yukawa-H4G}), (\ref{chilag}) and (\ref{mixing-TSS-MWT}); the full potential is given by adding (\ref{mixing-TSS-MWT}) to (\ref{higgs-potential}). The stationary condition, $0=\delta V/\delta\Phi |_{\langle\Phi_i\rangle }$, results in
\beq
0 &=& 
M^2_{11} + \lambda_b v^2_1 
+ \tan \beta \left[ -M^2_{12} + \frac{3}{2}\lambda'_b v^2_1 + \frac{1}{2}\lambda'_t v^2_2 \right]\,,
\\[1ex]
0 &=& 
M^2_{22} + \lambda_t v^2_2 
+ \cot \beta \left[ -M^2_{12} + \frac{1}{2}\lambda'_b v^2_1 + \frac{3}{2}\lambda'_t v^2_2 \right]\,.
\label{stationary-cond}
\eeq 

In addition to the vevs $v_1$ and $v_2$ of the top seesaw sector, we also have an electroweak symmetry breaking vev $v_{\rm TC}$ arising from MWT sector. These satisfy the constraint
$v_1^2+v_2^2+v_{\rm TC}^2=v_{\rm EW}^2$, where
$v_{\rm EW}=246 \, (\GeV)$. We define $\tan^2 \phi =v^2_{\rm TC}/(v^2_1 + v^2_2)$ ,
and then the above condition can be expressed as
\beq
v^2_1 + v^2_2 
=
v^2_{\rm EW} \cdot \cos^2 \phi
\label{vev-seesaw}
\,,
\eeq
%where $\tan^2 \phi =v^2_{\rm TC}/(v^2_1 + v^2_2)$ 
%and $\tan \beta = v_2/v_1$.

To analyse the physical mass spectrum of the higgs bosons in the present model, we simplify the parameter space by assuming that the states arising from the technicolor sector are heavy in comparison with the states arising from the topcolor sector. This means that we can obtain the physical spectrum by expanding Eq. (\ref{higgs-potential}) around $\vev{\Phi_i} = \left( 0\,, v_i/\sqrt{2}\right)^T$, and identifying the terms quadratic in the fields. 

The quadratic terms of the higgs boson fields are given by
\beq
{\cal L}^{\rm higgs}_m
=
-\frac{1}{2}(\pi^0_1 \,\,\, \pi^0_2) {\cal M}^2_{\pi} \bpm \pi^0_1 \\ \pi^0_2 \epm
-(\pi^+_1 \,\,\, \pi^+_2) {\cal M}^2_{\pi\pm} \bpm \pi^-_1 \\ \pi^-_2 \epm
-\frac{1}{2}(h^0_1 \,\,\, h^0_2) {\cal M}^2_{h} \bpm h^0_1 \\ h^0_2 \epm .
\eeq 
Here the mass matrix of CP-odd neutral higgs boson fields ($\pi^0_i$), including the neutral top-pion of the topcolor model, is
\beq
{\cal M}^2_\pi = 
\left[ M^2_{12} - \frac{1}{2} \lambda'_b v^2_1 - \frac{1}{2} \lambda'_t v^2_2 \right ]
\bpm \tan \beta & -1 \\ -1 & \cot \beta \epm\,.
\eeq
The mass matrix of charged higgs boson fields ($\pi^\pm_i$), which includes the charged top-pion, is
\beq
{\cal M}^2_{\pi \pm} = 
\left[ M^2_{12} + \frac{1}{2} \lambda_{tb} v_1v_2 - \frac{1}{2} \lambda'_b v^2_1 - \frac{1}{2} \lambda'_t v^2_2 \right]
\bpm \tan \beta & -1 \\ -1 & \cot \beta \epm\,,
\eeq
and the mass matrix of CP-even neutral higgs boson field ($h^0_i$) is
\beq
{\cal M}^2_h
\!\!\!&=&\!\!\!
\bpm 2 \lambda_1 v^2_1 & 0 \\ 0 & 2 \lambda_2 v^2_2 \epm 
+ M^2_{12} \bpm \tan \beta & -1 \\ -1 & \cot \beta \epm
\nonumber\\[1ex]
&&
+ \frac{1}{2}v_1v_2 
\bpm 
3\lambda'_b - \lambda'_t \tan^2\beta & 3\lambda'_b \cot \beta + 3 \lambda'_t \tan \beta
\\
3\lambda'_b \cot \beta + 3 \lambda'_t \tan \beta & 3\lambda'_t - \lambda'_b \cot^2\beta
\epm\,,
\eeq
where we define $\tan \beta \equiv v_2/v_1$. 
Among all higgs bosons, three fields 
%which we denote by $G^{0,\pm}$, 
are absorbed as the longitudinal modes of the electroweak gauge bosons. The remaining three CP-odd higgs boson fields, which we denote by $A^0\,,H^{\pm}$, and two CP-even higgs bosons fields appear as the physical higgs bosons in the low energy spectrum. 
The CP-odd physical higgs bosons and would-be Nambu-Goldstone bosons are related to the original fields of the seesaw sector as
\beq
\bpm G^0 \\ A^0 \epm
= 
\bpm \cos \beta & \sin \beta \\ - \sin \beta & \cos \beta \epm
\bpm \pi^0_1 \\ \pi^0_2 \epm
\quad , \quad
\bpm G^\pm \\ H^\pm \epm
= 
\bpm \cos \beta & \sin \beta \\ - \sin \beta & \cos \beta \epm
\bpm \pi^\pm_1 \\ \pi^\pm_2 \epm\,.
\eeq 

In this paper we consider a scenario where the physical low energy degrees of freedom arise dominantly from the top seesaw sector.  Hence, we consider the NGBs $\pi_{\rm TC}$, arising from the MWT sector, to contribute to the mixing so that the NGBs absorbed by $W^\pm,Z$ are represented by a linear combination of $G^i$ and $\pi_{\rm TC}^i$ as
\beq
G_{\rm absorbed}
=
 \cos \phi \left(\cos \beta  \pi_1
+\sin \beta  \right) \pi_2
+
\sin\phi \pi_{\rm TC}\,.
\eeq
%where  $\tan \beta = \frac{v_2}{v_1}$
%and $  \tan^2 \phi =\frac{v^2_{\rm TC}}{v^2_1 + v^2_2}$.
The orthogonal linear combination 
$ -\sin\phi \left(\cos \beta  \pi_1
+\sin \beta  \right) \pi_2
+
\cos\phi \pi_{\rm TC}$ we take to be heavy and decoupled.

The masses of the physical CP-odd higgs bosons are the non-zero eigenvalues of the matrices ${\cal M}^2_\pi$ and ${\cal M}^2_{\pi\pm}$,
\beq
M^2_{A} &=&
\frac{1}{\cos\beta \sin\beta}\left[ M^2_{12} - \frac{1}{2} \lambda'_b v^2_1 - \frac{1}{2} \lambda'_t v^2_2 \right ]
\,,\\[1ex]
M^2_{H\pm}&=&
M^2_A + \frac{1}{2} \lambda_{tb} \left[ v_1 \cos \beta + v_2 \sin \beta \right]^2\,.
\eeq
As to the CP-even higgs bosons sector, we define $\tan 2\alpha \equiv 2[{\cal M}^2_{h}]_{12}/([{\cal M}^2_{h}]_{11}-[{\cal M}^2_{h}]_{22})$ with $-\pi/2 \leq \alpha \leq 0$. Then the two mass eigenstates are represented by 
\beq
\bpm H^0 \\ h^0 \epm
= 
\bpm \cos \alpha & \sin \alpha \\ - \sin \alpha & \cos \alpha \epm
\bpm h^0_1 \\ h^0_2 \epm\,,
\eeq
and the corresponding mass eigenvalues are 
\beq
M^2_{H^0} 
&=& 
\frac{1}{2} 
\left[ 
[{\cal M}^2_{h}]_{11} + [{\cal M}^2_{h}]_{22} + 
\sqrt{([{\cal M}^2_{h}]_{11} - [{\cal M}^2_{h}]_{22})^2 + 4 [{\cal M}^2_{h}]^2_{12}}
 \right]\,
 \label{CP-even-heavy}
 \\[1ex]
M^2_{h^0} 
&=& 
\frac{1}{2} 
\left[ 
[{\cal M}^2_{h}]_{11} + [{\cal M}^2_{h}]_{22} - 
\sqrt{([{\cal M}^2_{h}]_{11} - [{\cal M}^2_{h}]_{22})^2 + 4 [{\cal M}^2_{h}]^2_{12}}
 \right]\,.
  \label{CP-even-light}
\eeq
It is convenient to rewrite $[{\cal M}^2_h]_{11,22,12}$ with the help of $M^2_A$ as 
\beq
[{\cal M}^2_h]_{11} 
&=& 
2\lambda_1v^2_1 + 2 \lambda'_b v_1 v_2 + M^2_A \sin^2\beta 
\,,
\label{CPeven-11}
\\[1ex]
[{\cal M}^2_h]_{22} 
&=& 
2\lambda_2v^2_2 + 2 \lambda'_t v_1 v_2 + M^2_A \cos^2\beta
\,,
\label{CPeven-22}
\\[1ex]
[{\cal M}^2_h]_{12} &=&
M^2_A + 2v_1 v_2 \left[ \lambda'_b \cot \beta +  \lambda'_t \tan \beta \right]
\label{CPeven-12}
\,.
\eeq
We note that as $v^2_{\rm EW}  = (246 \GeV)^2= v^2_{\rm TC} + v^2_1 + v^2_2$ in our model and  $\tan \phi \equiv v_{\rm TC}/\sqrt{v^2_1 +v^2_2}$,  with $\tan\phi =0$ we reproduce the results of the top quark seesaw model in \cite{He:2001fz}. 

\begin{figure}[htbp]
\begin{center}
\includegraphics[scale=0.75]{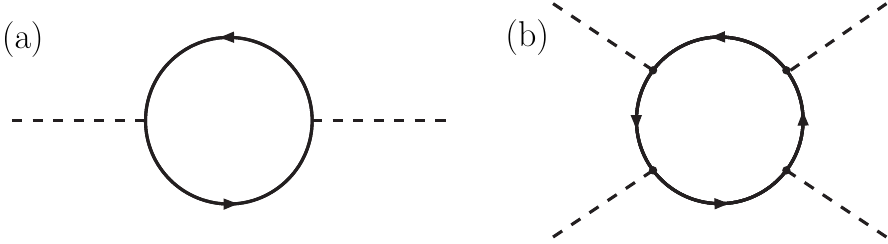} 
\caption{ 
Fermion contribution to the (a) two- and (b) four-point functions of  the higgs field. In these figures, solid line and dashed line correspond to fermion and higgs, respectively.
\label{BHL-bubble}}
\end{center}
\end{figure}

All these mass formulas are expressed in terms of the low energy effective theory fields and parameters. However,
by construction, the mass parameters ($M^2_{11,22,12}$) and quartic couplings ($\lambda_{b,t,tb}$,$\lambda'_{t,b}$) 
of the effective theory can be related to the parameters of the underlying ultraviolet theory by the direct computation of the 
two-point functions of the higgs shown in Fig.~\ref{BHL-bubble} (a), and four-point functions of the higgs shown in Fig.~
\ref{BHL-bubble} (b). 

Also the low energy fields $\Phi_i$ and couplings $y_i$ in the low energy Lagrangian are related to corresponding nonrenormalized fields  $\Phi^{(0)}_i$ and couplings $y_{i0}$ as $\Phi_{i}=Z^{1/2}_{\Phi_i}\Phi^{(0)}_i$ and $y_i = Z^{-1/2}_{\Phi_i} y_{i0}$. The wave function renormalization factors $Z_{\Phi_i}$  of the higgs are evaluated by a direct computation of the two-point function of the higgs in Fig.~\ref{BHL-bubble} (a) as 
\beq
Z_{\Phi_1} &=& 
\frac{3 y^2_{10}}{16 \pi^2} 
\left[ 
\sum^4_{\alpha=1} |D^L_{\alpha3}|^2|D^R_{\alpha4}|^2\ln\frac{\Lambda^2+m^2_{d\alpha}}{m^2_{d\alpha}}
+ 
\sum_{\alpha \neq \beta} |D^L_{\alpha3}|^2|D^R_{\beta4}|^2\ln\frac{\Lambda^2+m^2_{d4}}{m^2_{d4}}
\right] 
\,,\\
Z_{\Phi_2} &=& 
\frac{3 y^2_{20}}{16 \pi^2} 
\left[ 
\sum^4_{\alpha=1} |U^L_{\alpha3}|^2|U^R_{\alpha4}|^2\ln\frac{\Lambda^2+m^2_{u\alpha}}{m^2_{u\alpha}}
+ 
\sum_{\alpha\neq \beta} |U^L_{\alpha3}|^2|U^R_{\beta4}|^2\ln\frac{\Lambda^2+m^2_{u4}}{m^2_{u4}}
\right] 
\,.
\eeq
Here we have assumed $m_{di}\ll m_{d4} \,,\, m_{ui}\ll m_{u4}$ for $i=1,2,3$.  Consequently, the component fields in Eq.(\ref{higgs-paramet}) and their vacuum expectation values are
 also renormalized, e.g. $h^0_i = Z^{1/2}_{\Phi_i} h^0_{i0}$ and $v_i \equiv Z^{1/2}_{\Phi_i} \vev{h^0_{i0}}$.
% which correspond to the vacuum expectation values in the two higgs doublet models. 
In terms of these variables, the dynamical fermion masses $\Sigma_{U/D}$ are represented as $\Sigma_{U(D)}= y_{2(1)} v_{2(1)}/\sqrt{2}$, i.e.
\beq
v^2_1
\!\!\!&=&\!\!\!
\frac{3}{8\pi^2} \Sigma^2_D \left[ 
\sum^4_{\alpha=1} |D^L_{\alpha3}|^2|D^R_{\alpha4}|^2\ln\frac{\Lambda^2+m^2_{d\alpha}}{m^2_{d\alpha}}
+ 
\sum_{\alpha \neq \beta} |D^L_{\alpha3}|^2|D^R_{\beta4}|^2\ln\frac{\Lambda^2+m^2_{d4}}{m^2_{d4}}
\right] 
\label{decayconst-D}
\,,\\
v^2_2
\!\!\!&=&\!\!\!
\frac{3}{8\pi^2} \Sigma^2_U \left[ 
\sum^4_{\alpha=1} |U^L_{\alpha3}|^2|U^R_{\alpha4}|^2\ln\frac{\Lambda^2+m^2_{u\alpha}}{m^2_{u\alpha}}
+ 
\sum_{\alpha \neq \beta} |U^L_{\alpha3}|^2|U^R_{\beta4}|^2\ln\frac{\Lambda^2+m^2_{u4}}{m^2_{u4}}
\right]\,.
\label{decayconst-U}
\eeq
As stated earlier, we may safely ignore the topcolor instanton effects for the gap equations.

%
%This picture is done in the framework of linear sigma model, i.e. we assume that there are light composite higgs boson in a low energy effective theory based on the walking TC, however this is not consensus in the community, so it is fair to comment on a case in which there are not any light composite higgs (higgsless case : $h^0_{\rm TC}$ vanishes). For the higgsless case, not so complicate, we should rewrite $\Sigma_{\rm TC} \to v_{\rm TC}/\sqrt{2} \times \tilde{\Sigma}_{\rm TC}$ in Eq.(\ref{mixing-TSS-MWT}) where $\tilde{\Sigma}_{\rm TC} = \exp (2 i \tilde{\pi}_{\rm TC}/v_{\rm TC}) E_{SO(4)}$ is a non-linear field, so the first term in  Eq.(\ref{mixing-TSS-MWT}) vanishes and the resultant potential $V$ is the similar with the mixing term in  \cite{Chivukula:2011ag}.
The results of the calculation of the mass parameters and quadratic couplings in the low energy theory are collected in the Appendix \ref{matching}. To summarize the final results: 
The CP-odd higgs boson $A^0$ obtains a dynamical mass given by Eq. (\ref{dyn-CPodd-higgs}), the charged higgs boson $H^\pm$ obtains a dynamical mass given by Eq. (\ref{dyn-charged-higgs}) together with Eqs. (\ref{dyn-ch-11}),(\ref{dyn-ch-22}),(\ref{dyn-ch-12}). Finally, the CP-even higgs bosons $H^0,h^0\,\,(M_{H^0} > M_{h^0})$ have dynamical masses given by Eqs. (\ref{CP-even-heavy}),(\ref{CP-even-light}),(\ref{CPeven-11}),(\ref{CPeven-22}),(\ref{CPeven-12}) together with Eqs. (\ref{relation-prime12}),(\ref{dyn-CPeven-1}),(\ref{dyn-CPeven-2}). 
These masses, as well as the fourth generation quarks masses, $m_{t',b'}$, are determined for given values the parameters $\tan \beta$, $\tan\phi$, $\xi$, $\Lambda$ with the criticality condition $g_{U/D} > 1$.

%%%%%%%%%%%
%%%%%%%%%%%
%%%%%%%%%%%
%%%%%%%%%%%

%%%%%%%%%%%%%%%%%%%%%%%%%%%%%%%%%%%%%%%%%%
%%%%%%%%%%%%%%%%%%%%%%%%%%%%%%%%%%%%%%%%%%
%%%%%%%%%%%%%%%%%%%%%%%%%%%%%%%%%%%%%%%%%%
%
\section{Oblique corrections: $S$ and $T$ parameters}
%%%%%%%%%%%%%%%%%%%%%%%%%%%%%%%%%%%%%%%%%%
%%%%%%%%%%%%%%%%%%%%%%%%%%%%%%%%%%%%%%%%%%
%
%%%%%%%%%%%%%%%%%%%%%%%%%%%%%%%%%%%%%%%%%%
%
%\subsection{$S$ and $T$ parameters in the present model}
\label{ST-ours}
%
%%%%%%%%%%%%%%%%%%%%%%%%%%%%%%%%%%%%%%%%%%

Representing the self energy of the electroweak gauge bosons as $\Pi^{\mu \nu}_{XY}(q^2) = g^{\mu\nu}  \Pi_{XY}(q^2) + (q^\mu q^\nu\text{-term})$ where $XY=+-,33,3Y$, the Peskin-Takeuchi $S,T$-parameters \cite{Peskin:1990zt} are given by
\beq
S \!\!&\equiv&\!\! -16 \pi \frac{\Pi_{3Y}(M^2_Z) - \Pi_{3Y}(0) }{M^2_Z}
\label{PT-S}
\,,\\
T \!\!&\equiv&\!\! \frac{4\pi}{M^2_Zs^2_W c^2_W} \left[ \Pi_{+-}(0) - \Pi_{33}(0) \right]\,.
\label{PT-T}
\eeq
In the present model the contributions from sectors beyond the SM are  
\beq
S \!\!&=&\!\!  S_{N,E} + S_{q4} + S_{\rm TC} + S_{\rm higgs} + S_{G',Z'} +\Delta_S \,,\\
T \!\!&=&\!\!  T_{N,E} + T_{q4} + T_{\rm TC} + T_{\rm higgs} + T_{G',Z'} + \Delta_T\,,
\eeq
where the first five terms on the right hand side correspond to, from left to right, the contributions of the fourth leptons, the contributions of the fourth quarks, the contributions of the technicolor sector, the higgs contributions and coloron and $Z'$ contirbutions. The last term $\Delta_{S,T}$ is
\beq
\Delta_{S,T} = -\Delta_{S,T}(x^{\rm ref}_h) + \Delta_{S,T}(x^{\rm ours}_h).
\eeq
This is the contribution from the light CP-even higgs boson ($h^0$). The SM-higgs contribution with $m^{\rm ref}_h$ is subtracted from the SM values $S_{\rm SM} = T_{\rm SM} =0$. Here $x_h \equiv m^2_h/M^2_Z$. In the present paper we use $\Delta_{S,T}$ as defined in Appendix C of \cite{Hagiwara:1994pw} (and denoted by $H_S$ ($H_T$) for $S$ ($T$) there).

\subsubsection{Fourth generation fermions contribution}
\label{ST-fouthgeneration}
The leading contributions to the $S,T$ parameters from the fourth generation leptons are \cite{Lavoura:1992np,He:2001tp}
\beq
S_{N,E}&=&\frac{1}{2\pi} \left[ \psi_\nu (x_N) + \psi_e (x_E) \right]\,,
\\[1ex]
T_{N,E}&=&\frac{1}{16 \pi s^2_W c^2_W} \theta_+(x_N,x_E)\,,
\eeq
where we do not consider a mixing between the fourth generation leptons and the other three generation leptons, and we take the fourth generation leptons heavy, $x_N \gg x_{\nu i}$ and $x_E \gg x_{ei}$ where $x_i \equiv m^2_i/M^2_W$, $i=1,2,3$. On the other hand, since we consider the mixing for quark sectors, following \cite{Lavoura:1992np}, the one-loop contributions from the fourth generation quarks sector are given by 
\beq
S_{q4}
=
\frac{3}{2\pi}
\left[ 
\begin{aligned}
& \left( 1 - |U^{(L)}_{44}|^2 \right) \psi_u (x_{t'})
- \sum^3_{\alpha=1} |U^{(L)}_{4\alpha}|^2 \psi_u (x_{u\alpha})
\\[1ex]
&
+ \left( 1 - |D^{(L)}_{44}|^2 \right) \psi_d (x_{b'})
- \sum^3_{\alpha=1} |D^{(L)}_{4\alpha}|^2 \psi_d (x_{d\alpha})
\\[1ex]
&
- \sum^3_{\alpha=1}|U^{(L)}_{4\alpha}|^2 |U^{(L)}_{44}|^2 \chi_+(x_{u\alpha},x_{t'}) 
- \sum_{\alpha<\beta}\left(|U^{(L)}_{4\alpha}|^2 |U^{(L)}_{4\beta}|^2 -1\right) \chi_+(x_{u\alpha},x_{u\beta}) 
\\[1ex]
&
- \sum^3_{\alpha=1}|D^{(L)}_{4\alpha}|^2 |D^{(L)}_{44}|^2 \chi_+(x_{d\alpha},x_{b'}) 
- \sum_{\alpha<\beta}\left(|D^{(L)}_{4\alpha}|^2 |D^{(L)}_{4\beta}|^2 -1\right) \chi_+(x_{d\alpha},x_{d\beta}) 
\end{aligned}
\right]\,,
\label{S-vf}
\eeq

\beq
T_{q4}
= 
\frac{3}{16 \pi s^2_W c^2_W} 
\left[ 
\begin{aligned}
&
\sum^3_{\alpha=1} |U^{(L)}_{\alpha4}|^2 |D^{(L)}_{\alpha4}|^2 \theta_+(x_{t'},x_{b'})
+ \sum^3_{\alpha,\beta,\gamma=1} \left( |U^{(L)}_{\alpha\beta}|^2 |D^{(L)}_{\alpha\gamma}|^2 -1 \right)\theta_+(x_{u\beta},x_{d\gamma})
\\[1ex]
&
- \sum^3_{\alpha=1} |U^{(L)}_{4\alpha}|^2 |U^{(L)}_{44}|^2 \theta_+(x_{u\alpha},x_{t'})
- \sum_{\alpha<\beta} \left(|U^{(L)}_{4\alpha}|^2 |U^{(L)}_{4\beta}|^2 -1\right)\theta_+(x_{u\alpha},x_{u\beta})
\\[1ex]
&
- \sum^3_{\alpha=1} |D^{(L)}_{4\alpha}|^2 |D^{(L)}_{44}|^2 \theta_+(x_{d\alpha},x_{b'})
- \sum_{\alpha<\beta} \left(|D^{(L)}_{4\alpha}|^2 |D^{(L)}_{4\beta}|^2 -1\right)\theta_+(x_{d\alpha},x_{d\beta})
\end{aligned}
\right]\,,
\label{T-vf}
\eeq
where $x_{u\alpha} \equiv m^2_{u\alpha}/M^2_W$ and $x_{d\alpha} \equiv m^2_{d\alpha}/M^2_W$ and the functions $\psi_{\nu,e,u,d}(x)$ are given by 
\beq
\psi_{+1/2}(Y,x) 
\!\!\!&=&\!\!\!
 \left( \frac{8Y}{3} + 2 \right) x -\frac{2Y}{3} \ln x + \frac{(4Y+3)x + 2Y}{6} f(x,x)
\,,\\[1ex]
\psi_{-1/2}(Y,x) 
\!\!\!&=&\!\!\! 
\left( \frac{-8Y}{3} + 2 \right) x +\frac{2Y}{3} \ln x + \frac{(-4Y+3)x -2Y}{6}  f(x,x)
\,,\\[1ex]
\psi_\nu(x) \!\!\!&=&\!\!\! \psi_{+1/2}(-1/2,x)
\quad,\quad
\psi_e(x) = \psi_{-1/2}(-1/2,x)\,,
\\[1ex]
\psi_u(x) \!\!\!&=&\!\!\! \psi_{+1/2}(1/6,x)
\quad , \quad 
\psi_d(x) = \psi_{-1/2}(1/6,x)\,,
\eeq
and the functions $\chi_+,\theta_+,f$ are given in Appendix \ref{integral}.

%%%%%%%%%%%%%%%%%%%%%%%%%%%%%%%%%%%%%%%%%%
%
\subsubsection{Technicolor contribution}
\label{ST-TC}
%
%%%%%%%%%%%%%%%%%%%%%%%%%%%%%%%%%%%%%%%%%%
The technicolor contribution is due to the minimal walking technicolor model \cite{Sannino:2004qp,Dietrich:2005jn}. This is a walking technicolor theory whose strong coupling gauge theory dynamics are nonperturbative. However, the perturbative one-loop formulae are often used to estimate the leading contributions to $S,T$-parameters, and we take this estimate
as a guide also in this paper. Therefore, we obtain
\beq
S_{\rm TC} \!\!\!&=&\!\!\! 
\frac{N_D d[R]}{2\pi} \left[ \psi_{+1/2}(Y,x_{\cal T}) +  \psi_{-1/2}(Y,x_{\cal B})  \right]
\,,\label{S-TC} \\[1ex]
T_{\rm TC} \!\!\!&=&\!\!\!
\frac{N_D d[R]}{16 \pi s^2_W c^2_W} \theta_+(x_{\cal T},x_{\cal B})\,,
\label{T-TC} 
\eeq
where $x_{\cal T} \equiv m^2_{\cal T}/M^2_Z$ and $x_{\cal B} \equiv m^2_{\cal B}/M^2_Z$,
$R$ denotes the representation of the technifermions under the technicolor gauge group, $d[R]$ is a dimension of the representation $R$ and $N_D$ is the number of $SU(2)_L$ doublets.
In the present model we have $R =\Ysymm$, $d[\Ysymm] = 3$ and $N_D = 1$ with $Y=1/6$. When we take $m_{\cal T} =  m_{\cal B} \gg M_Z$, Eqs. (\ref{S-TC}) and (\ref{T-TC}) are 
\beq
S_{\rm TC} = \frac{1}{2\pi} 
\quad \text{and} \quad 
T_{\rm TC} = 0\,.
\eeq
Moreover, the $S$-parameter indicates a decreasing tendency thanks to the walking dynamics \cite{Appelquist:1991is,Sundrum:1991rf,Harada:1994ni,Appelquist:1998xf,Ignjatovic:1999ch,Harada:2005ru,Kurachi:2006mu,Haba:2008nz}. The value of $S_{\rm TC}$ including an effect of the walking dynamics is estimated as $ \sim S_{\rm TC} \times 0.7 \simeq 0.1$ \cite{Sundrum:1991rf}. Due to the nonperturbative nature of these estimates we will allow for a broader range of values, $S=0.1,\dots 0.3$, in the study of the electroweak precision constraints for the present model in section~\ref{numerical-results}.

%%%%%%%%%%%%%%%%%%%%%%%%%%%%%%%%%%%%%%%%%%
%
\subsubsection{Higgs contributions}
\label{ST-dynamicalhiggs}
%
%%%%%%%%%%%%%%%%%%%%%%%%%%%%%%%%%%%%%%%%%%

The leading contributions to the $S,T$ parameters from the dynamical higgs sector are given by \cite{He:2001fz,He:2001tp} 
\beq
S_{\rm higgs}
\!\!\!&=&\!\!\!
\frac{1}{\pi M^2_Z} 
\left[ 
\begin{aligned}
&
\sin^2(\beta - \alpha) \bar{\cal F}'_2(M^2_Z,m^2_H,m^2_A) - \bar{\cal F}'_2(M^2_Z,m^2_{H^\pm},m^2_{H^\pm}) 
\\[1ex]
&
+
\cos^2(\beta - \alpha) 
\left\{ \bar{\cal F}'_2(M^2_Z,m^2_h,m^2_A) + \bar{\cal F}'_2(M^2_Z,m^2_Z,m^2_H) - \bar{\cal F}'_2(M^2_Z,m^2_Z,m^2_h) \right\}
\\[1ex]
&
+ \cos^2(\beta - \alpha) M^2_Z \left\{ {\cal F}'_0(M^2_Z,m^2_Z,m^2_H) - {\cal F}'_0(M^2_Z,m^2_Z,m^2_h) \right\}
\end{aligned}
\right]\,,\nonumber\\
\eeq
and 
\beq
T_{\rm higgs}
\!\!\!&=&\!\!\!
\frac{1}{16\pi s^2_W c^2_W}
\left[
\begin{aligned}
&
\theta_+(x_{H^\pm},x_A) + \sin^2(\beta-\alpha) \left\{ \theta_+(x_{H^\pm},x_H) - \theta_+(x_A,x_H) \right\}
\\[1ex]
&\hspace*{11.7ex}
+ \cos^2(\beta-\alpha) \left\{ \theta_+(x_{H^\pm},x_h) - \theta_+(x_A,x_h) \right\}
\\[1ex]
&
+ \cos^2(\beta-\alpha) 
\left\{ \begin{aligned}
&
\theta_+(x_W,x_H) - \theta_+(x_W,x_h) + 4 \bar{\cal F}'_0(M^2_W,m^2_H,m^2_h)
\\[1ex]
&
-\theta_+(x_Z,x_H) + \theta_+(x_Z,x_h)  - 4 \bar{\cal F}'_0 (M^2_Z,m^2_H,m^2_h)
\end{aligned}\right\}
\end{aligned}
\right]\,,
\nonumber\\
\eeq
where
\beq
\bar{\cal F}'_2(q^2,m^2,M^2)
\!\!\!&=&\!\!\!
\frac{ q^2}{2} \left[ \frac{1}{6}\ln q^2 - {\cal F}_2 (q^2,m^2,M^2) + {\cal F}_2 (0,m^2,M^2) \right]
\,,\\[1ex]
\bar{\cal F}'_0(m^2,M^2_1,M^2_2)
\!\!\!&=&\!\!\!
\frac{m^2}{M^2_Z}
\left[ {\cal F}_0(0,m^2,M^2_1)- {\cal F}_0(0,m^2,M^2_2) \right]
\,,
\eeq
and the functions ${\cal F}_{0,2},{\cal F}'_0,\theta_+$ are given in Appendix \ref{integral}.

%%%%%%%%%%%%%%%%%%%%%%%%%%%%%%%%%%%%%%%%%%
%
\subsubsection{Coloron and $Z'$ contributions}
\label{ST-bNJL}
%
%%%%%%%%%%%%%%%%%%%%%%%%%%%%%%%%%%%%%%%%%%

\begin{figure}[htbp]
\begin{center}
\begin{flushleft} {(a)}\end{flushleft}
\includegraphics[scale=0.75]{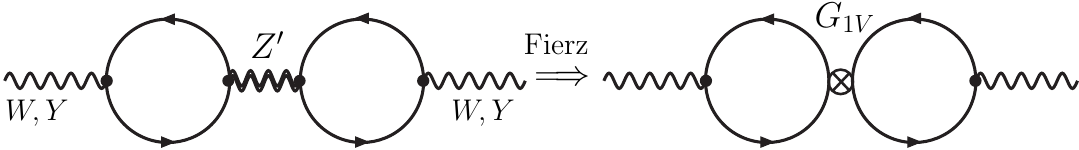} \\[1ex]
\begin{flushleft} (b)\end{flushleft}
\includegraphics[scale=0.75]{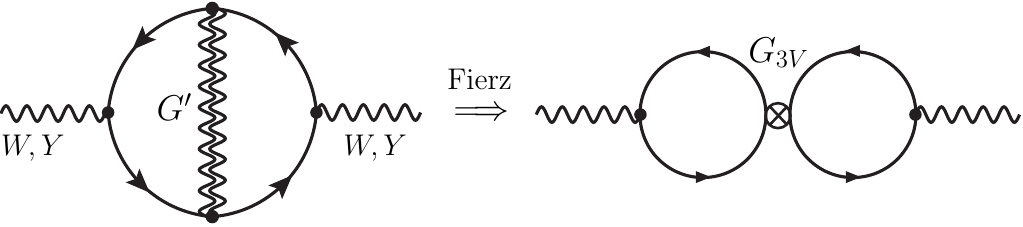} 
\caption{ 
(a) $Z'$ and (b) $G'$ contributions to the self energy of the electroweak gauge bosons. 
The diagrams on the left show the leading $Z',G'$ contribution which corresponds to the two-loop diagrams. The diagrams on the right show these diagrams after replacing the massive vector boson exchange by vector four fermion interactions and Fierz rearranging \cite{Chivukula:1995dc}. As a result, the diagrams correspond to the product of two one-loop diagrams. \label{vacuum-polarization-4f}}
\end{center}
\end{figure}

We should also take into account the contributions to the EWPT observables from the massive gauge bosons $Z'$ and $G'$. 
% The former appears after the $U(1)_{Y1} \times U(1)_{Y2}$ breaks to $U(1)_Y$ and the latter appears after the $SU(3)_1 \times SU(3)_2$ breaks to $SU(3)_{\rm QCD}$. 
As one can see from Eqs. (\ref{4f-3}),(\ref{4f-1}), the leading contributions from $Z'$ and $G'$ to the self energy of the electroweak gauge bosons are given by left diagrams in Fig.~\ref{vacuum-polarization-4f}. 
Although these leading contributions are two-loop diagrams, they become a product of two one-loop diagrams after a Fierz rearrangement of the four fermion interaction corresponding to the exchange of a heavy vector boson \cite{Chivukula:1995dc}.%; see Eq.(\ref{Fierz-LL},\ref{Fierz-RR}) . 

Among the four fermion interactions given by Eq. (\ref{4f-scalar})  and Eq. (\ref{4f-vector}), the scalar four fermion interactions to the vacuum polarizations of the electroweak gauge bosons appear as $\propto q^\mu q^\nu$, where $q$ is the momentum of the electroweak gauge bosons. Hence, these do not contribute to the electroweak precision parameters. Due to this fact, we do not show $G_{1S,3S}$ in Fig.~\ref{vacuum-polarization-4f}. 

We will ignore the contribution from $SU(3)_2$ and $U(1)_{Y2}$ sectors because these are proportional to $\alpha^2_{\rm QCD}/\kappa_3, \alpha^2_Y/\kappa_1 \ll 1$, respectively, as shown in Eq.(\ref{4f-3},\ref{4f-1}). Moreover, even though the third generation leptons have $U(1)_{Y1}$ charge (see Table ~\ref{particle-model}) their masses are smaller than the masses of $u^{(3,4)},d^{(3,4)}$ quarks, and hence we ignore also their contributions.

\begin{figure}[htbp]
\begin{center}
\includegraphics[scale=1]{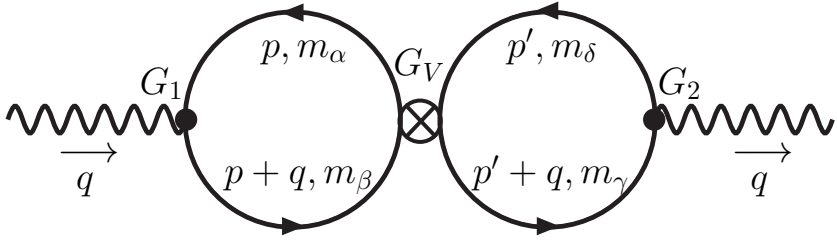} 
\caption{ 
For calculations. 
\label{vacuum-polarization-4f-fc}}
\end{center}
\end{figure}%

Let us now compute diagrams in Fig. \ref{vacuum-polarization-4f-fc} for a general case where $G_{1(2)} = g_{L1(2)} P_L +g_{R1(2)} P_R  $. The coupling $G_V$ ($\otimes$ in figure) is a vector four fermion coupling which has a form as $[\cdots \gamma_\mu P_{L(R)} \cdots ] \times [\cdots \gamma^\mu P_{L(R)} \cdots] $ as shown in Eq.(\ref{4f-vector}), so we represent $G_V$ as $G^{L \otimes L}_V$ and  $G^{R \otimes R}_V$, respectively. For a case with $L \otimes L$, Fig. \ref{vacuum-polarization-4f-fc} is calculated as
\beq
\Pi^L(q^2)
\!\!\!&=&\!\!\!
-\frac{G^{L \otimes L}_V}{2} \left( \frac{3}{8\pi^2} \right)^2  
\left[ g_{R1} m_\alpha m_\beta \bar{\cal F}_0(q^2,m^2_\alpha,m^2_\beta) 
- g_{L1} \bar{\cal F}_1(q^2,m^2_\alpha,m^2_\beta)\right]
\nonumber\\[1ex]
&& \hspace*{15ex}
\times
\left[ g_{R2} m_\delta m_\gamma \bar{\cal F}_0(q^2,m^2_\delta,m^2_\gamma) 
- g_{L2} \bar{\cal F}_1(q^2,m^2_\delta,m^2_\gamma)\right]
\,,
\label{vp-LL}
\eeq
and for a case with $R \otimes R$, Fig. \ref{vacuum-polarization-4f-fc} is calculated as
\beq
\Pi^R(q^2)
\!\!\!&=&\!\!\!
-\frac{G^{R \otimes R}_V}{2} \left( \frac{3}{8\pi^2} \right)^2  
\left[ g_{L1} m_\alpha m_\beta \bar{\cal F}_0(q^2,m^2_\alpha,m^2_\beta) 
- g_{R1} \bar{\cal F}_1(q^2,m^2_\alpha,m^2_\beta)\right]
\nonumber\\[1ex]
&& \hspace*{15ex}
\times
\left[ g_{L2} m_\delta m_\gamma \bar{\cal F}_0(q^2,m^2_\delta,m^2_\gamma) 
- g_{R2} \bar{\cal F}_1(q^2,m^2_\delta,m^2_\gamma)\right]
\,,
\label{vp-RR}
\eeq
where 
\beq
\bar{\cal F}_0(q^2,m^2,M^2)
\!\!\!&=&\!\!\!
\ln \frac{q^2}{\Lambda^2} + {\cal F}_0(q^2,m^2,M^2) 
\,,\\[1ex]
\bar{\cal F}_1(q^2,m^2,M^2)
\!\!\!&=&\!\!\!
q^2 \left[ \frac{2-3(x+y)}{6} \ln \frac{q^2}{\Lambda^2} + {\cal F}_1(q^2,m^2,M^2) \right]
\,,
\eeq
and the functions ${\cal F}_{0,1}$ are given in Appendix \ref{integral}. 
\begin{table}[h]
\begin{center}
{
\renewcommand\arraystretch{2.5}
\begin{tabular}{| c  || c | c |}
\hline
four fermion operators & $-i G^{L\otimes L}_V$ or $-i G^{R\otimes R}_V$
\\
\hline
\parbox[c][7.5ex][c]{0ex}{}
$[\bar{u}^{(\alpha)} \gamma^\mu P_L u^{(\beta)}] [\bar{u}^{(\gamma)} \gamma_\mu P_L u^{(\delta)}]$ 
& $-i\left[G_{3V} + \dfrac{1}{36} G_{1V} \right] \cdot 
\left( U^{L*}_{3\alpha}\, U^L_{3\beta} \right) \cdot \left(U^{L*}_{3\gamma}\,  U^L_{3\delta} \right)$
\\ \hline
\parbox[c][7.5ex][c]{0ex}{}
$[\bar{u}^{(\alpha)} \gamma^\mu P_R u^{(\beta)}] [\bar{u}^{(\gamma)} \gamma_\mu P_R u^{(\delta)}]$ 
& $-i\left[G_{3V} + \dfrac{4}{9} G_{1V} \right] \cdot \left( U^{R*}_{4\alpha}\, U^R_{4\beta} \right) \cdot \left(U^{R*}_{4\gamma}\,  U^R_{4\delta} \right)$
\\ \hline \hline
\parbox[c][7.5ex][c]{0ex}{}
$[\bar{d}^{(\alpha)} \gamma^\mu P_L d^{(\beta)}] [\bar{d}^{(\gamma)} \gamma_\mu P_L d^{(\delta)}]$ 
& $-i\left[G_{3V} + \dfrac{1}{36} G_{1V} \right] \cdot \left( D^{L*}_{3\alpha}\, D^L_{3\beta} \right) \cdot \left(D^{L*}_{3\gamma}\,  D^L_{3\delta} \right)$
\\ \hline 
\parbox[c][10ex][c]{0ex}{}
$[\bar{d}^{(\alpha)} \gamma^\mu P_R d^{(\beta)}] [\bar{d}^{(\gamma)} \gamma_\mu P_R d^{(\delta)}]$ 
& $-i\left[G_{3V} + \dfrac{1}{9} G_{1V} \right] \cdot \left( D^{R*}_{4\alpha}\, D^R_{4\beta} \right) \cdot \left(D^{R*}_{4\gamma}\,  D^R_{4\delta} \right)$
\\ \hline \hline
\parbox[c][10ex][c]{0ex}{}
$[\bar{u}^{(\alpha)} \gamma^\mu P_L d^{(\beta)}] [\bar{d}^{(\gamma)} \gamma_\mu P_L u^{(\delta)}]$ 
& $-2i \left[G_{3V} + \dfrac{1}{36} G_{1V} \right] \cdot \left( U^{L*}_{3\alpha}\, D^L_{3\beta} \right) \cdot \left(D^{L*}_{3\gamma}\,  U^L_{3\delta} \right)$
\\
\hline
\parbox[c][10ex][c]{0ex}{}
$[\bar{u}^{(\alpha)} \gamma^\mu P_R d^{(\beta)}] [\bar{d}^{(\gamma)} \gamma_\mu P_R u^{(\delta)}]$ 
& $-2i \left[G_{3V} - \dfrac{2}{9} G_{1V} \right] \cdot \left( U^{R*}_{4\alpha}\, D^R_{4\beta} \right) \cdot \left(D^{R*}_{4\gamma}\,  U^R_{4\delta} \right)$
\\
\hline
\end{tabular}
}
\caption{The vector four quarks operators in terms of mass eigenstates for quarks which is derived from Eq.(\ref{4f-vector}) with Eq.(\ref{quark-mixing}). The fermion with index $\alpha=1,\dots,4$ has mass $m_\alpha$ in Fig.~\ref{vacuum-polarization-4f-fc}.} 
\label{vertex-4fV}
\end{center}
\end{table}%
In the present model, one can express each $G_V$ in terms of quark mass eigenstates from Eqs. (\ref{4f-vector}) and (\ref{quark-mixing}). The resulting vertex coefficients are given in Table ~\ref{vertex-4fV}. 
 
 By substituting $G_{V}$ in Table.~\ref{vertex-4fV} and $g_{L,R}$ in  Table.~\ref{vertex-ffEW} into Eqs.  (\ref{vp-LL}) and (\ref{vp-RR}), and summing over all quark generations in loops, we obtain each $\Pi(q^2)$.  For example, let us consider
 \beq
 \Pi_{3Y}(q^2) = \Pi^L_{3Y}(q^2) + \Pi^R_{3Y}(q^2),\qquad \Pi^{L,R}_{3Y}(q^2) = \Pi^{L,R}_{3Y}(q^2)\left.\right|_{u} + \Pi^{L,R}_{3Y}(q^2)\left.\right|_{d}. 
 \eeq
 For $\Pi^L_{3Y}(q^2)\left.\right|_{u}$, one can read each $g_{L,R}$ from Table ~\ref{vertex-ffEW} as
 \beq
 g_{L1} &=& (1/2) {\cal U}^L_{\beta \alpha},\quad g_{R1}=0,\quad g_{L2} = -(1/2) \, {\cal U}^L_{\delta \gamma}+ (2/3) \, \delta_{\delta \gamma},\nonumber \\
 g_{R2} &=& (2/3)\, \delta_{\delta \gamma},\quad G^{L \otimes L}_V = [G_{3V} + (1/36) G_{1V}] \cdot \left( U^{L*}_{3\alpha}\, U^L_{3\beta} \, U^{L*}_{3\gamma}\,  U^L_{3\delta} \right). 
 \eeq
 Thus we obtain
\beq
\Pi^L_{3Y}(q^2) \left.\right|_u
\!\!\!&=&\!\!\!
+\frac{1}{4} \left[G_{3V} + \dfrac{1}{36} G_{1V} \right] \left( \frac{3}{8\pi^2}\right)^2 
 \label{vp-3Y-LL-u}  \\[1ex]
&& \!\!\!\!
\times
\sum_{\alpha,\beta,\gamma,\delta} 
 \left[ U^{L*}_{3\alpha}\, U^L_{3\beta} \, U^{L*}_{3\gamma}\,  U^L_{3\delta} \right] \cdot
{\cal U}^L_{\beta\alpha} \bar{\cal F}_1(q^2,m^2_\alpha,m^2_\beta)
%\nonumber\\[1ex]
%&& \hspace*{15ex}
%\times
\left\{ \begin{aligned}
&
\frac{2}{3} m^2_\gamma \bar{\cal F}_0(q^2,m^2_\gamma,m^2_\gamma) 
\\[1ex]
&
- \left( -\frac{1}{2} {\cal U}^L_{\delta \gamma}+ \frac{2}{3} \delta_{\delta \gamma}\right) \bar{\cal F}_1(q^2,m^2_\delta,m^2_\gamma)
\end{aligned}\right\}
\,.\nonumber
\eeq
For all other $\Pi(q^2)$s, it is easy to obtain a similar representation as Eq. (\ref{vp-3Y-LL-u}). By substituting each result for $\Pi(q^2)$ into Eqs. (\ref{PT-S}) and (\ref{PT-T}), we obtain $S_{G',Z'}$ and $T_{G',Z'}$;  we do not present their formulas explicitly.

%%%%%%%%%%%%%%%%%%%%%%%%%%%%%%%%%%%%%%%%%%
%%%%%%%%%%%%%%%%%%%%%%%%%%%%%%%%%%%%%%%%%%
%%%%%%%%%%%%%%%%%%%%%%%%%%%%%%%%%%%%%%%%%%
%
%
%
\section{Numerical results}
\label{numerical-results}
%
%
%%%%%%%%%%%%%%%%%%%%%%%%%%%%%%%%%%%%%%%%%%
%%%%%%%%%%%%%%%%%%%%%%%%%%%%%%%%%%%%%%%%%%
In this section we perform the numerical calculations for the present model by using the results derived in previous sections. In this paper, we fix $g^{(34)}_{U,D}$ in Eq. (\ref{dless-4f}) as 
\beq
g^{(34)}_U = g^{(34)}_D+ \frac{1}{9} \kappa_1 = 1.2
\quad \text{and} \quad
\kappa_1 = 0.5 \,,
\label{assumption-coupling}
\eeq
which satisfies the criticality condition $g^{(34)}_{U,D} > g_{\crit} =1$ and the Landau pole constraint for $\Lambda_L/\Lambda =10$ as one can see from Fig.~\ref{gaptriangle-model}. 
The value of $g_{U,D}^{(34)}$ is dictated to lie above but near the critical value by the requirement to reproduce correct masses for the third generation quarks. The parameter $\kappa_1$ is constrained to lie in a narrow range $\kappa_1<0.6$ by the Landau pole constraint, and the results are not heavily affected by its value.

Moreover we assume that the quark mixing matrices Eq.(\ref{quark-mixing}) reflect the seesaw mechanism for the third and fourth generations. This implies that the third and fourth generation mixing will dominate these matrices. Thus, at the leading order, these matrices are \cite{He:2001fz} $( 0 < \theta^{u,d}_{L,R} < \pi/2)$, 
\beq
U^L_{\alpha\beta}\simeq
\bpm 
1 & 0 & 0 & 0 \\
0 & 1 & 0 & 0 \\
0 & 0 & \cos \theta^u_L & \sin \theta^u_L \\
0 & 0 & -\sin \theta^u_L & \cos \theta^u_L
\epm
\quad &,& \quad 
U^R_{\alpha\beta}
\simeq
\bpm 
1 & 0 & 0 & 0 \\
0 & 1 & 0 & 0 \\
0 & 0 & -\cos \theta^u_R & \sin \theta^u_R \\
0 & 0 & \sin \theta^u_R & \cos \theta^u_R
\epm\,,
\,\label{seesaw-mixing-u}
%\\[2ex]
%D^L_{\alpha\beta} =U^L_{\alpha\beta}\left.\right|_{u \to d}
%\quad \hspace*{17ex} &,& \quad
%D^R_{\alpha\beta} = U^R_{\alpha\beta}\left.\right|_{u \to d}\,.
%\label{seesaw-mixing-d}
\eeq
and similarly for $D^L_{\alpha\beta}$ and $D^R_{\alpha\beta}$ by replacing $u\rightarrow d$ in the above definitions.
With these matrices, we diagonalize the mass terms in which the dynamical mass term $\Sigma_U \overline{U^{3}_L} U^{(4)}_R + \Sigma_D \overline{D^{3}_L} D^{(4)}_R$, together with its hermitian conjugate, is combined with Eq.(\ref{baremass-model2}), and we identify each eigenvalues as the third generation quark masses $m_t$, $m_b$ and the fourth generations quarks masses $m_{t'} (> m_t)$,and $m_{b'}(>m_b)$. The eigenvalues $m_{t'}$ and $m_{b'}$ are related to given $m_t$ and $m_b$ as
\beq
m_{t'} = m_t \cdot \left[ \cot \theta^u_L \cot \theta^u_R \right]
\quad , \quad
m_{b'} = m_b \cdot \left[ \cot \theta^d_L \cot \theta^d_R \right]\,,
\label{t'b'-mass}
\eeq  
where $m_t = 172.9\, \GeV$ and $m_b = 4.2\, \GeV$. 
We note that the above mixing matrices are only a leading approximation, and a more refined structures may be inferred from phenomenology. For example, in order to explain the Tevatron anomaly in the top quark forward backward asymmetry \cite{Aaltonen:2011kc} by the model of this type, one may need a mixing between the light quarks and the heavy third generations quarks \cite{Bai:2011ed}. 
% we should take into account all components of $U^{L/R}$ and $D^{L/R}$ with %keeping the reflection of the seesaw mechanism for full phenomenology. 
However, since these effects are subleading and we do not aim to explain the Tevatron anomaly quantitatively,  we will use Eqs. (\ref{seesaw-mixing-u}) %,\ref{seesaw-mixing-d}) 
as the quark mixing matrices in our analysis here. 

Under these assumptions, and for given $\Lambda = M_{G',Z'}$ and topcolor instanton parameter $\xi$ in section.~\ref{LEEFT-higgs}, we now proceed to calculate numerically
\begin{itemize}
\item masses of the fourth generation quarks $t',b'$ by solving the gap equations Eqs. (\ref{gapeq-U-af}) and (\ref{gapeq-D-af}),
\item dynamical higgs masses derived in section ~\ref{LEEFT-higgs},
\item the Peskin-Takeuchi $S,T$-parameters as given in section~\ref{ST-ours}. 
\end{itemize}

%%%%%%%%%%%%%%%%%%%%%%%%%%%%%%%%%%%%%%%%%%
%%%%%%%%%%%%%%%%%%%%%%%%%%%%%%%%%%%%%%%%%%
%
\subsection{Dynamical results for mass spectrums of fourth generation quarks and  higgs}
%
%%%%%%%%%%%%%%%%%%%%%%%%%%%%%%%%%%%%%%%%%%
When we substitute $m_{t',b'}$ given by Eq. (\ref{t'b'-mass}) into the gap equations (\ref{gapeq-U-af}) and (\ref{gapeq-D-af}), the resulting equations depend only on $\theta^{u,d}_{L,R}$ and $m_{t,b}$; the couplings $g^{(34)}_{U,D}$ take the values given in Eq.(\ref{assumption-coupling}). Since the dynamical symmetry breaking derived from $g^{(34)}_{U,D} > g_{\rm crit}$ gives only a part of the electroweak gauge boson masses, we should solve the resultant gap equations under a conditions in which the decay constants Eqs.(\ref{decayconst-D}) and (\ref{decayconst-U}) together with $v_{\rm TC}$ satisfy $v_1^2+v_2^2+v_{\rm TC}^2=v_{\rm EW}^2$, where
$v_{\rm EW}=246 \, (\GeV)$. This condition is more conveniently written as
\beq
v^2_1 + v^2_2 
=
v^2_{\rm EW} \cdot \cos^2 \phi
\label{vev-seesaw}
\,,
\eeq
where $\tan^2 \phi =v^2_{\rm TC}/(v^2_1 + v^2_2)$.

In this paper, we assume the walking technicolor sector has characteristics of a low scale technicolor model \cite{Eichten:1979ah,Lane:1989ej,Lane:2009ct,Delgado:2010bb}, meaning that we set $\tan \phi \leq 1$. In the limit $\tan \phi = 0$, we obtain the original top quark seesaw model. Moreover, $T_{G',Z'}$ in the topcolor model becomes large and {\it positive} at $\Lambda \simeq {\cal O}(\TeV)$ \cite{Chivukula:1995dc}, on the other hand, $T_{4q}$ in the top seesaw model becomes large and {\it negative} at $\Lambda \simeq {\cal O}(\TeV)$ \cite{He:2001fz}, so we can expect a cancellation between $T_{G',Z'}$ and $T_{4q}$ at $\Lambda \simeq {\cal O}(\TeV)$ if we take $\tan \beta =v_2/v_1\leq1$. 
%Thus, in this paper we take $\tan \phi$ and $\tan \beta$ as 
%\beq
%0 \leq \tan \phi \leq 1\quad , \quad \tan \beta =1\,,
%\eeq
Specifically, we will consider two special cases $\tan \phi = 0, 1$ with $\tan\beta=1$ in the following.

\begin{figure}[htbp]
\begin{center}
\includegraphics[scale=0.6]{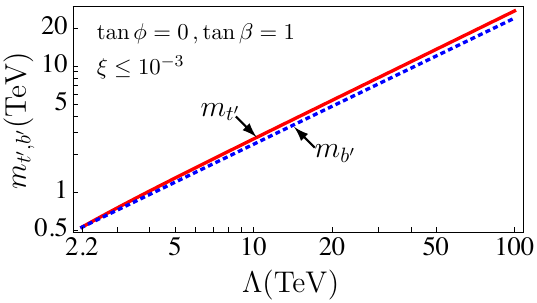} \quad
\includegraphics[scale=0.6]{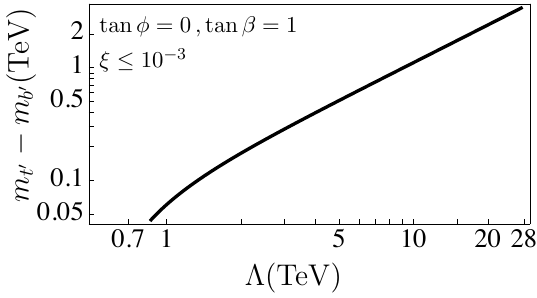} 
%\\[2ex]
%\includegraphics[scale=0.75]{mtpmbp-Lambda-b1a1o4.pdf} \quad
%\includegraphics[scale=0.75]{mtpmbpdiff-b1a1o4.pdf} 
%\\[2ex]
%\includegraphics[scale=0.75]{mtpmbp-Lambda-b1a1o2.pdf} \quad
%\includegraphics[scale=0.75]{mtpmbpdiff-b1a1o2.pdf} 
\\[2ex]
\includegraphics[scale=0.6]{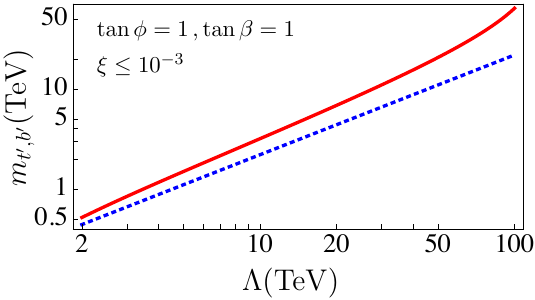} \quad\,
\includegraphics[scale=0.6]{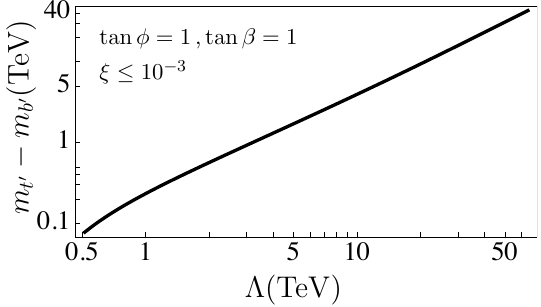} 
\caption{ 
The dynamical results for fourth generations quarks. 
The left and right panels show $m_{t'} (\TeV)$ (red/solid) and $m_{b'} (\TeV)$ (blue/dotted) and $m_{t'}-m_{b'} (\TeV)$ respectively. From top to bottom, $(\tan \phi,\tan \beta) = (0,1), (1,1)$ with $\xi \leq 10^{-3}$. The most top figures correspond to the top quark seesaw model in \cite{He:2001fz} with $g = 1.2$ and $\kappa_1=0$.
\label{4th-quark-mass}}
\end{center}
\end{figure}

\begin{figure}[htbp]
\begin{center}
\begin{tabular}{cc}
{
\begin{minipage}[t]{0.5\textwidth} 
%\begin{flushleft} (a) \end{flushleft} \vspace*{-5ex}
\includegraphics[scale=0.5]{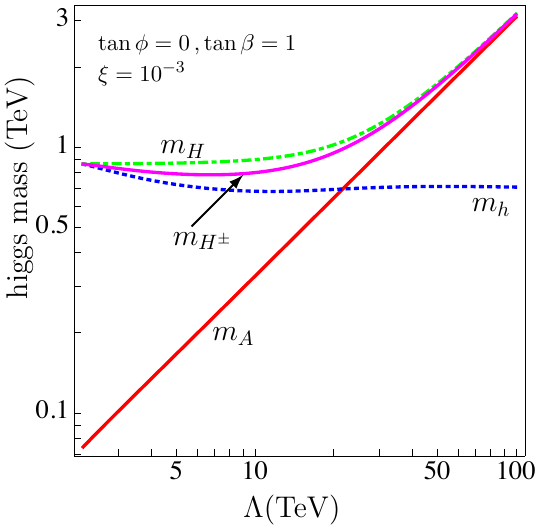} 
\vspace*{5ex}
\end{minipage}
}%{
%\begin{minipage}[t]{0.5\textwidth}
%\begin{flushleft} (b) \end{flushleft} \vspace*{-5ex}
%\includegraphics[scale=0.75]{mhiggs-b1a1o4.pdf}
%\vspace*{5ex}
%\end{minipage} 
%}\\
%{
%\begin{minipage}[t]{0.5\textwidth}
%\begin{flushleft} (c) \end{flushleft} \vspace*{-5ex}
%\includegraphics[scale=0.75]{mhiggs-b1a1o2.pdf} 
%\end{minipage}
%}
{
\begin{minipage}[t]{0.5\textwidth}
%\begin{flushleft} (b) \end{flushleft} \vspace*{-5ex}
\includegraphics[scale=0.5]{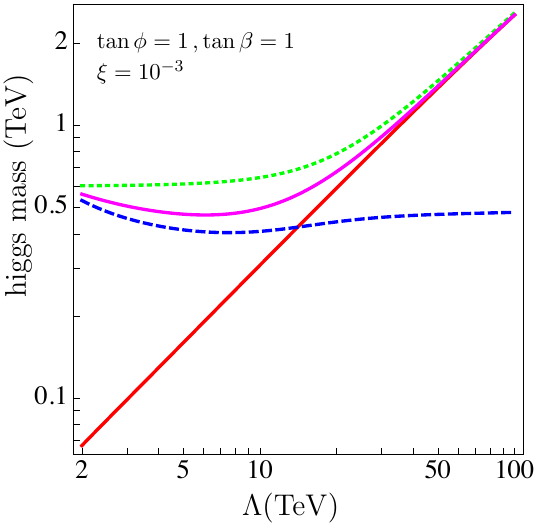}
\end{minipage}
}
\end{tabular}
\caption{ 
The dynamical results for {\it higgs} mass for $2 \,\TeV \leq \Lambda \leq 100 \,\TeV$. 
In all figures, $m_{A^0}$, $m_{H^\pm}$, $m_{h^0}$, $m_{H^0}$ are represented as solid/red, solid/magenta, dotted/blue and dash-dotted/green line, respectively. 
These panels are $(\tan \phi,\tan \beta) = (0,1)$ (left) and $ (1,1)$ (right). The left panel corresponds to the top quark seesaw model in \cite{He:2001fz} with $g = 1.2$, $\kappa_1=0$ and $\xi = 10^{-3}$.
\label{higgs-mass}}
\end{center}
\end{figure}

In Fig.\ref{4th-quark-mass} we show the numerical results for $m_{t',b'}$ and $m_{t'}-m_{b'}$,  and in Fig.\ref{higgs-mass} for $m_{A^0}$, $m_{H^\pm}$, $m_{h^0}$ and $m_{H^0}$  when $2 \, \TeV \leq \Lambda \leq 100 \, \TeV$, $\tan \beta = 1$ and $\tan \phi = 0,1$. We reproduce the results of \cite{He:2001fz} at the limit $\kappa_1=0$ and $\tan \phi = 0$ in our model. As $\tan \phi \to 0$, mass splitting $m_{t'}-m_{b'}$ becomes small at $\Lambda \simeq {\cal O}(\TeV)$ and, due to the restoration of the custodial symmetry, $m_{H^\pm} \simeq m_{h^0} \simeq m_{H^0}$ at $\Lambda \simeq {\cal O}(\TeV)$. We find that, as $\tan \phi \to 1$, $\kappa_1$ tends to influence the mass splitting even if the cut-off scale is several TeV. From Fig.\ref{higgs-mass}, we can see easily that dynamical higgs mass in the present model can be smaller than in the top quark seesaw model. In the present model, the light CP-even neutral higgs mass $M_{h^0}$ is $400 - 500 \, \GeV$ which is smaller than $M_{h^0} = 700 - 900 \,\GeV$ obtained in the top quark seesaw model. This is so due to the existence of the technicolor sector which allows $v^2_1 + v^2_2$ to be small in the limit $\tan\phi\rightarrow 1$.

%%%%%%%%%%%%%%%%%%%%%%%%%%%%%%%%%%%%%%%%%%
%%%%%%%%%%%%%%%%%%%%%%%%%%%%%%%%%%%%%%%%%%
%
\subsection{The Peskin-Takeuchi $S,T$-parameters}
%
%%%%%%%%%%%%%%%%%%%%%%%%%%%%%%%%%%%%%%%%%%

%\subsection{$S,T$ parameters by Gfitter}
%\label{EW-Gfitter}
%
%%%%%%%%%%%%%%%%%%%%%%%%%%%%%%%%%%%%%%%%%%

To consider the electroweak precision tests (EWPT), we need the constraints for Peskin-Takeuchi $S,T,U$-parameters \cite{Peskin:1990zt} from the electroweak precision data. %For this purpose, we use 13 observables which are $M_W$,$\sin^2 \theta^l_{\rm eff}$, $\Gamma_Z$,$\sigma^0_{\rm had}$, $R^0_l$, $R^0_c$, $R^0_b$, $A^l$, $A^c$, $A^b$, $A^{0,l}_{\rm FB}$, $A^{0,c}_{\rm FB}$, $A^{0,b}_{\rm FB}$ with lepton universality. Their experimental values and SM theoretical values are shown in Table.\ref{for-EWPT-constraints}. In this paper, we consider the representation in \cite{Peskin:1990zt} for these observables as a function of the parameters $S$ and $T$,  while setting $U=0$. We use the experimental values and SM values for these EWPT observables as in the Gfitter results \cite{Flacher:2008zq}.  As a result we obtain
In this paper we use values in \cite{Nakamura:2010zzi} as
\beq
S = 0.03  \pm 0.09
\quad , \quad
T = 0.07 \pm 0.08\,,
\label{ST-Gfitter}
\eeq
and a correlation $\rho_{ST} = 0.87$ for a reference higgs mass $m^{\rm ref}_h = 117\, \GeV$. Note that these values for $S,T$ parameters differ with respect to \cite{Ludwig:2010vk}. This is because \cite{Nakamura:2010zzi} fix $U=0$, i.e. for a two parameter fit, while the estimation in \cite{Ludwig:2010vk} is for a three parameter fit. A fact that values in \cite{Nakamura:2010zzi} are slightly larger than the results for three parameters fit \cite{Nakamura:2010zzi}.

\begin{figure}[htbp]
\begin{center}
\includegraphics[scale=0.6]{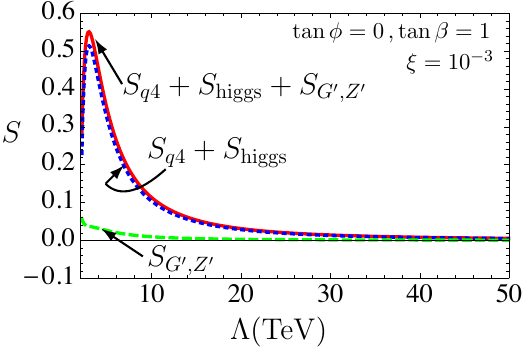} \quad
\includegraphics[scale=0.6]{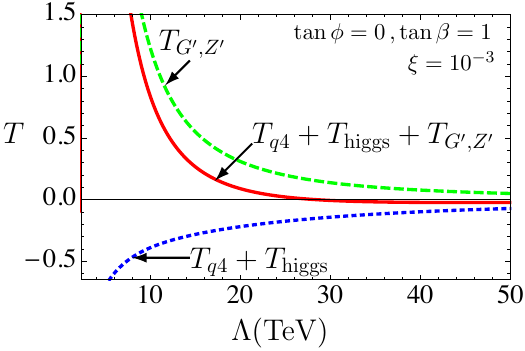} 
%\\[2ex]
%\includegraphics[scale=0.75]{Stot-b1a1o4.pdf} \quad
%\includegraphics[scale=0.75]{Ttot-b1a1o4.pdf} 
%\\[2ex]
%\includegraphics[scale=0.75]{Stot-b1a1o2.pdf} \quad
%\includegraphics[scale=0.75]{Ttot-b1a1o2.pdf} 
\\[2ex]
\includegraphics[scale=0.6]{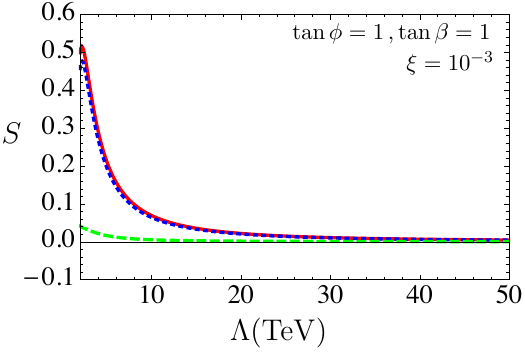} \quad
\includegraphics[scale=0.6]{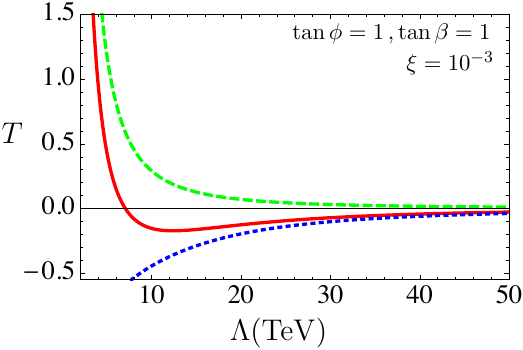} 
\caption{ 
The Peskin-Takeuchi $S$-parameter (left panels) and $T$-parameter (right panels) coming from the fourth generation quarks in section.\ref{ST-fouthgeneration}, the dynamical {\it higgs} in section.\ref{ST-dynamicalhiggs} and $G',Z'$ in in section.\ref{ST-bNJL} for $2 \,\TeV \leq \Lambda \leq 50 \,\TeV$. In all figures, the dashed/green, the dotted/blue and the solid/red curves correspond to $S({\rm or \,} T)_{G',Z'}$, $S({\rm or \,} T)_{4q} + S({\rm or \,} T)_{higgs}$ and $S({\rm or \,} T)_{4q} + S({\rm or \,} T)_{higgs} + S({\rm or \,} T)_{G',Z'}$, respectively. Top and Bottom panels correspond to $(\tan \phi,\tan \beta) = (0,1), (1,1)$. The top figures correspond to the top quark seesaw model in \cite{He:2001fz} with $g = 1.2$ and $\kappa_1=0$.
\label{ST-4thquark-higgs-V4f}}
\end{center}
\end{figure}

\begin{figure}[htbp]
\begin{center}
\begin{tabular}{cc}
{
\begin{minipage}[t]{0.5\textwidth}
\begin{flushleft} (a) \end{flushleft} \vspace*{-5ex}
\includegraphics[scale=0.62]{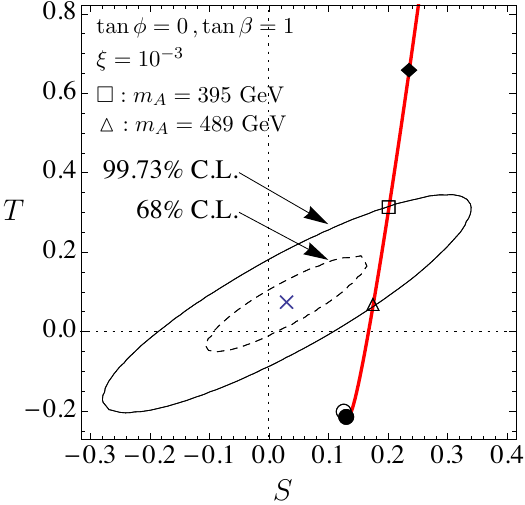} 
\vspace*{2ex}
\end{minipage}
}
{
\begin{minipage}[t]{0.5\textwidth}
\begin{flushleft} (b) \end{flushleft} \vspace*{-5ex}
\includegraphics[scale=0.6]{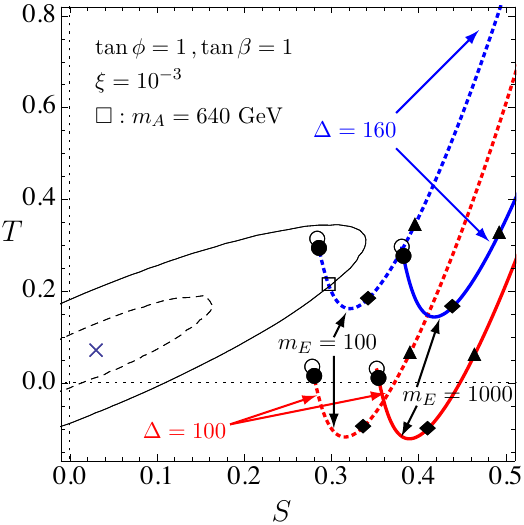} 
%\vspace*{5ex}
\end{minipage}
}
\\
%{
%\begin{minipage}[t]{0.5\textwidth}
%\begin{flushleft} (b) \end{flushleft} \vspace*{-2ex}
%\includegraphics[scale=0.75]{ST-b1a14-4n.pdf} 
%\vspace*{5ex}
%\end{minipage}
%}
%\\[2ex]
%{
%\begin{minipage}[t]{0.5\textwidth}
%\begin{flushleft} (c) \end{flushleft} \vspace*{-2ex}
%\includegraphics[scale=0.75]{ST-b1a12-4n.pdf}
%\vspace*{5ex}
%\end{minipage}
%}
{
\begin{minipage}[t]{0.5\textwidth}
\begin{flushleft} (c) \end{flushleft} \vspace*{-5ex}
\includegraphics[scale=0.6]{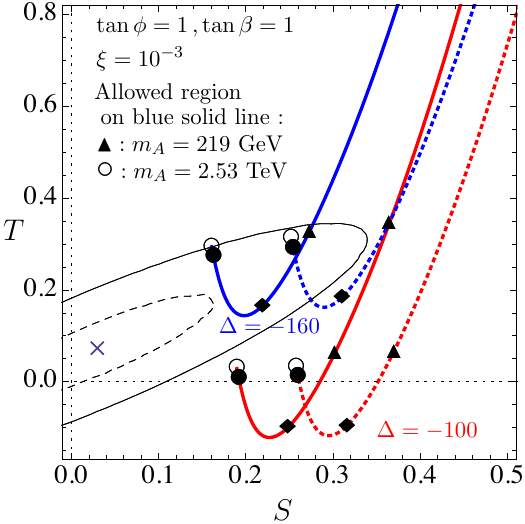} 
%\vspace*{5ex}
\end{minipage}
}
{
\begin{minipage}[t]{0.5\textwidth}
\begin{flushleft} (d) \end{flushleft} \vspace*{-5ex}
\includegraphics[scale=0.6]{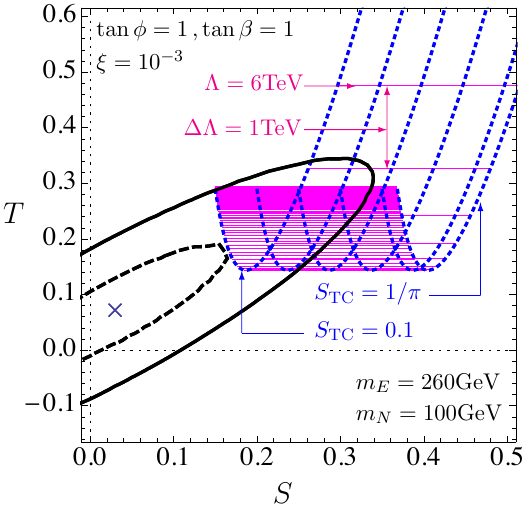} 
%\vspace*{5ex}
\end{minipage}
}
\end{tabular}
\caption[]{
The Peskin-Takeuchi $S,T$-parameters in the present model with experimentally allowed contour at $99.73\% (68 \%) \cl$, corresponding to the solid (dashed) ellipsis. Panel (a) corresponds to $(\tan \phi,\tan \beta) = (0,1)$  and (b,c) correspond to $(\tan \phi,\tan \beta) = (1,1)$. In panel (b) the upper (lower) set of two curves corresponds to $\Delta \equiv m_N - m_E = 160\,(100) \,\GeV$  with $m_E = 100\, \GeV$ (solid curve) and $1000\,\GeV$ (dotted curve). In panel (c) the upper (lower) curves correspond to $\Delta= -160\,(-100) \,\GeV$  with $m_N = 100\, \GeV$ (solid curve) and $1000\,\GeV$ (dotted curve). On each curve the symbols \blacktriangles,\blackdiamond,\blackcircles and \whitecircles correspond to $\Lambda = 7\,\TeV$, $10\,\TeV$, $50\,\TeV$ and $100\,\TeV$, respectively. In panel (a) {\scriptsize $\square$} and {\scriptsize$\triangle$} correspond to $(\Lambda (\TeV), m_A (\GeV))= (12, 395)$ and $(15,489)$. In panel (b) {\scriptsize$\square$} on the dotted line with $\Delta = 170$ corresponds to $(\Lambda (\TeV), m_A (\GeV))= (15.9, 474)$, and in panel (c) \blacktriangles and \whitecircles on the solid line with $\Delta = -160$ correspond to $(\Lambda (\TeV), m_A (\GeV))= (7, 219)$ and $(\Lambda (\TeV), m_A (\TeV))= (100, 2.5)$, respectively. In panel (d), for a case with $(m_N (\GeV),m_E (\GeV)) = (100,260)$, the dotted lines correspond to $S_{\rm TC} = 0.1,0.15,0.2,0.25,0.3,1/\pi$ from right to left. The horizontal (magenta) lines show various $\Lambda$ from $6 \TeV$ to $100 \TeV$ at intervals of $\Delta \Lambda (\TeV) = 1$.
\label{ST-present-model}}
\end{center}
\end{figure}

In our model, there are several sources contributing to the precision parameters, and to understand various effects we consider different contributions separately before presenting the full analysis. In Fig.\ref{ST-4thquark-higgs-V4f}, we show $S_{4q} + S_{higgs} + S_{G',Z'}$ for $2 \,\TeV \leq \Lambda \leq 50\,\TeV$, and also the corresponding contribution to the $T$ parameter. In the evaluation of these quantities we use the dynamical results obtained in previous sections. In addition, for comparison, we also show $S_{4q} + S_{higgs}$ and  $S_{G',Z'}$ (and the corresponding contributions to the $T$ parameters) in Fig.\ref{ST-4thquark-higgs-V4f}. From these figures we see, as we had expected, that the massive gauge boson contribution cancels with a large contribution from the dynamical seesaw sectors, i.e. vector-like quarks and the resultant higgs contributions for $T$ parameter. As to the $S$-parameter, the massive gauge bosons contributions are smaller than the contributions from the dynamical seesaw sectors. Unfortunately, a total value of the $S$ parameter coming from the topcolor sectors is too large for a low $\Lambda$; a result already obtained in \cite{He:2001fz}. But, in the present model, there are the fourth family leptons which can make a negative contribution to the $S$ parameter for a suitable mass difference between $N^{(4)}$ and $E^{(4)}$. In Fig.\ref{ST-present-model}, we show total $S,T$-parameters in the present model. In this figure, the solid (dashed) ellipsis,  obtained from Eq(\ref{ST-Gfitter}) 
%in section \ref{EW-Gfitter}, 
show the experimentally allowed contour at $99.73\% (68 \%) \cl$. Panel (a) corresponds to $(\tan \phi,\tan \beta) = (0,1)$, i.e. the top quark seesaw model in  \cite{He:2001fz} with $g = 1.2$ and $\kappa_1=0$. In panels (b) and (c) we have set $(\tan \phi,\tan \beta) = (1,1)$, and the curves show how the model results are affected by the masses of fourth generation leptons. In panel (b) the upper (lower) set of two curves correspond to $\Delta \equiv m_N - m_E = 160\, (100) \,\GeV$ with $m_E = 100\, \GeV$ (solid curve) and $1000\,\GeV$ (dotted curve), while in panel (c)  the upper (lower) curves correspond to $\Delta = -160\, (-100) \,\GeV$ with $m_N = 100\, \GeV$ (solid curve) and $1000\,\GeV$ (dotted curve).
On each curve, the symbols \blacktriangles, \blackdiamond, \blackcircles and \whitecircles correspond, respectively, to $\Lambda = 7\,\TeV$, $10\,\TeV$, $50\,\TeV$ and $100\,\TeV$. In panel (a), the symbols {\scriptsize $\square$} and {\scriptsize$\triangle$} correspond to $(\Lambda (\TeV), m_A (\GeV))= (12, 395)$ and $(15,489)$. In panel (b) the symbol {\scriptsize$\square$} on the dotted line corresponds to $(\Lambda (\TeV), m_A (\GeV))= (21.9, 640)$. In panel (c) the symbol\blacktriangles and \whitecircles on the solid line correspond to $(\Lambda (\TeV), m_A (\GeV))= (7, 219)$ and $(\Lambda (\TeV), m_A (\TeV))= (100, 2.5)$, respectively. In (a),(b),(c), the value of $S_{\rm TC}$ is fixed with $0.1$. In order to see the dependence with $S_{\rm TC}$, in panel (d), we show various case with varying $S_{\rm TC}$ from $0.1$ to $1/\pi$ with $(m_N (\GeV), m_E (\GeV)) = (260,100)$.The most left dotted curve in (d) corresponds to the upper solid line in panel (c).

We can see from these figures that if we take $\tan \phi =1$ and $\Delta = m_N - m_E = -170 \,\GeV$ with $m_N = 100\,\GeV - 1\,\TeV$ we can take the cutoff as low as $\Lambda = 7\,\TeV$ and the CP-odd higgs $A^0$ mass is around $200\,\GeV$ with $m_{t'} \simeq 2.1\,\TeV$ and $m_{b'} \simeq 1.5 \,\TeV$. Such a light CP-even higgs is not allowed in the top quark seesaw model with $\tan \beta =1$. In fact, we can see from Fig.\ref{ST-present-model}(a), together with results from Fig.\ref{higgs-mass}(a), that the allowed mass region is given by $395 \,\GeV \lessim m_{A^0} \lessim 490\,\GeV$ in a case of the top quark seesaw model with $\tan \beta =1$ and $\xi = 10^{-3}$.

We note that relative to the accuracy of the precision data, the largest uncertainty is in the $S$-parameter of the Technicolor sector. However, as we have shown, our model features several other sources to $S$ and $T$, which are perturbatively under control. Hence, the analysis presented in this section will be useful in indicating how the model parameters can 
be used to assess the viability of the model if the contributions from the technicolor sector can be determined more precisely by using, say, lattice simulations.

%%%%%%%%%%%%%%%%%%%
\subsection{$R_b$ constraint}
In the present model, we should take into consideration a constraint for $Zb\bar{b}$-vertex since, in general, this constraint is stronger than the constraint on the $S,T$-plane in the top quark seesaw model \cite{He:2001fz}. In this paper it is enough to consider only the quark sector 
although the dynamical higgs sector contributes to the $Zb\bar{b}$-vertex at the one-loop level. This is so since the latter contribution is large when $H^\pm$ is light as $m_{H^\pm} \simeq 200 \GeV$ \cite{Haber:1999zh}, which is not the case in the scenarios we consider; see e.g. Fig.\ref{higgs-mass}. Therefore it is enough to consider only the tree level contribution to the $Zb\bar{b}$-vertex in the present paper. 

\begin{figure}[htbp]
\begin{center}
\begin{tabular}{cc}
{
\begin{minipage}[t]{0.5\textwidth}
%\begin{flushleft} (a) \end{flushleft} \vspace*{-5ex}
\includegraphics[scale=0.62]{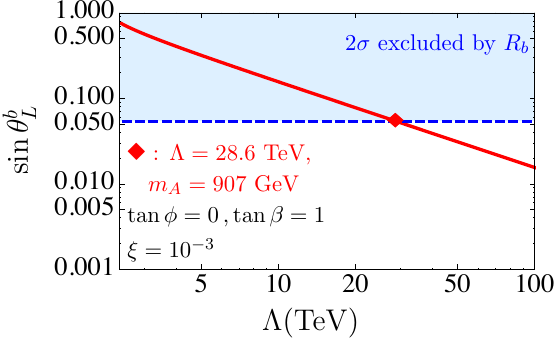} 
%\vspace*{2ex}
\end{minipage}
}
%{
%\begin{minipage}[t]{0.5\textwidth}
%\begin{flushleft} (b) \end{flushleft} \vspace*{-2ex}
%\includegraphics[scale=0.75]{ST-b1a14-4n.pdf} 
%\vspace*{5ex}
%\end{minipage}
%}
%\\[2ex]
%{
%\begin{minipage}[t]{0.5\textwidth}
%\begin{flushleft} (c) \end{flushleft} \vspace*{-2ex}
%\includegraphics[scale=0.75]{ST-b1a12-4n.pdf}
%\vspace*{5ex}
%\end{minipage}
%}
{
\begin{minipage}[t]{0.5\textwidth}
%\begin{flushleft} (b) \end{flushleft} \vspace*{-5ex}
\includegraphics[scale=0.6]{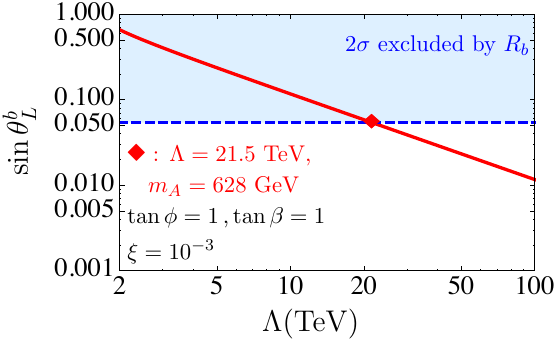} 
%\vspace*{5ex}
\end{minipage}
}
\end{tabular}
\caption[]{
$R_b$ constraint on $(\Lambda,\sin \theta^b_L)$-plane for the present model with parameters
 $(\tan \phi,\tan \beta) = (0,1)$ (left) and $ (1,1)$ (right) and $\xi = 10^{-3}$.
The red/solid line corresponds to $\sin \theta^b_L (\Lambda)$, which is obtained as dynamical solution, the blue dashed line corresponds to $2\sigma$ $R_b$ constraint and the shaded region above blue dashed line is $2\sigma$ excluded region. The diamond in both figures shows the $2\sigma$ lower bound for $\Lambda$, i.e. the lower bound is $(\Lambda (\TeV), m_A (\GeV))= (28.6, 907)$ (left) and $(21.5,628)$ (right). The left panel corresponds to the top quark seesaw model in \cite{He:2001fz} with $g = 1.2$ and $\kappa_1=0$.
\label{Rb-present-model}}
\end{center}
\end{figure}

The experimental value of $R_b$ \cite{Baak:2011ze} is
\beq
R^{\rm exp}_b  \equiv \frac{\Gamma(Z \to \bar{b}b)}{\Gamma(Z \to {\rm had})}=0.21629 \pm 0.00066 .
\label{Rb-exp}
\eeq
It is convenient to divide $R_b= R^{\rm SM}_b + \Delta R_b$,
where $R^{\rm SM}_b$ is presented as \cite{Baak:2011ze}
\beq
R^{\rm SM}_b = 0.21578 ^{+ 0.00005}_{-0.00008} \,.
\eeq
The quantity $\Delta R_b$ then encapsulates the contribution from the new physics (NP), and is represented as 
\beq
\Delta R_b = 2 R^{\rm SM}_b (1-R^{\rm SM}_b) 
{\rm Re} \left[
\frac{g^b_L \left[ \delta g^b_L\right]_{\rm NP} + g^b_R \left[ \delta g^b_R\right]_{\rm NP}}{(g^b_L)^2 + (g^b_R)^2}
\right].
\label{dRb}
\eeq
The experimental data constrains its value as
\beq
\Delta R_b = 0.00051 \pm 0.00066\,.
\label{Rb-NPconstraint}
\eeq
Eq.(\ref{dRb}) is derived straightforwardly from \cite{Hollik:1988ii} and $g^b_{L,R}$ is the SM tree level value given by
\beq
g^b_L = -\frac{1}{2} + \frac{1}{3}\sin^2\theta_W\,,\quad
g^b_R = \frac{1}{3} \sin^2\theta_W\,,
\eeq
In the present model, a contribution to the $Zb\bar{b}$-vertex is given by \cite{He:2001fz}
\beq
\delta g^b_L = \frac{e}{2s_Wc_W}\sin^2 \theta^b_L
\quad, \quad
\delta g^b_R =0\,,
\eeq
where $e = \sqrt{4\pi\alpha}$ with $\alpha \simeq 1/128$. We show the resulting constraint in the $(\Lambda,\sin \theta^b_L)$-plane for the present model in Fig.\ref{Rb-present-model}. The solid line shows the dynamical solution $\sin\theta^b_L(\Lambda)$ obtained from the model, while the dashed horizontal line corresponds to the $2\sigma$ constraint on $R_b$; the shaded region above this line is excluded. We can read off from Fig.\ref{Rb-present-model} that the lower bound is $(\Lambda, m_A )= (28.6 \TeV, 907 \GeV)$ for $(\tan \beta, \tan \phi) = (1,0)$  and $(\Lambda , m_A )=(21.5 \TeV,628\GeV)$ for $(\tan \beta, \tan \phi) = (1,1)$. Comparing these results with the EWPT parameters results as shown in Fig.\ref{ST-present-model} illustrates the fact that $R_b$ constraint is stronger than the EWPT constraints in this type of models \cite{He:2001fz}. We find that in the case with $(\tan \beta, \tan \phi) = (1,0)$ there are no overlap between EWPT parameters constraint and the $R_b$ constraint. However, in the case with $(\tan \beta, \tan \phi) = (1,1)$ there are allowed overlap: e.g. $628\, \GeV \leq m_A \leq 2.53 \,\TeV$ corresponding to $21.5\, \TeV \leq \Lambda \leq 100\,\TeV$ thanks to the existence of the fourth family leptons with with $(m_N,m_E) = (100\, \GeV, 260\, \GeV)$. 

%We should comment on this $R_b$ constraint results.
Note that these results may be sensitive to the contribution from the vector mesons in the MWT sector as shown in \cite{Fukano:2011aq,Fukano:2011iw}. However, these results
were derived for effective theory defined on the coset space $SU(2)_L \times SU(2)_R/SU(2)_V$ 
while $G/H$ in the MWT is $SU(4)/SO(4)$. 

%%%%%%%%%%%%%%%%%%%
\subsection{LHC phenomenology}
In this section we consider the model in light of present and future LHC data. The mass of the color octet vector boson is given by $M_{G'}=\Lambda$, and at $95 \% \cl$  $M_{G'} > 3.32 \,\TeV$ \cite{Aad:2011fq}. Hence, the 4th family quarks are constrained by $m_{t'} > 900 \,\GeV$ and $m_{b'} > 700 \,\GeV$ for $\tan\beta=\tan\phi=1$, as we can read off from Fig.\ref{4th-quark-mass}. These constraints for the 4th family quark masses are consistent with $m_{t'} > 422 \,\GeV$ at $95 \% \cl$ \cite{CMS-23Aug2011}, $m_{t'} > 450 \,\GeV$ at $95 \% \cl$ \cite{CMS-22Jul2011} and $m_{b'} > 361 \,\GeV$ at $95 \% \cl$ \cite{Chatrchyan:2011em}. On the other hand, the lightest PNGBs of the model may be discovered in the LHC data when considering the SM higgs boson production via the gluon fusion and decay into two photons\cite{ATLAS-13Dec2011,CMS-13Dec2011}. We consider next this possibility.

\begin{figure}[h]
\begin{center}
\begin{tabular}{cc}
{
\begin{minipage}[t]{0.5\textwidth}
\begin{flushleft} (a) \end{flushleft} \vspace*{-5ex}
\includegraphics[scale=0.6]{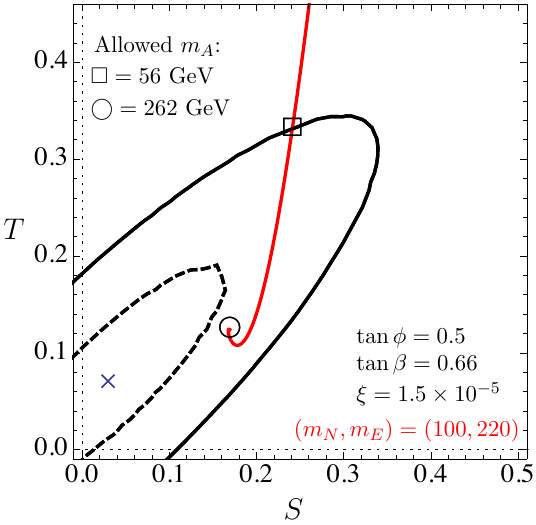} 
%\vspace*{5ex}
\end{minipage}
}
{
\begin{minipage}[t]{0.5\textwidth}
\begin{flushleft} (b) \end{flushleft} \vspace*{-5ex}
\includegraphics[scale=0.6]{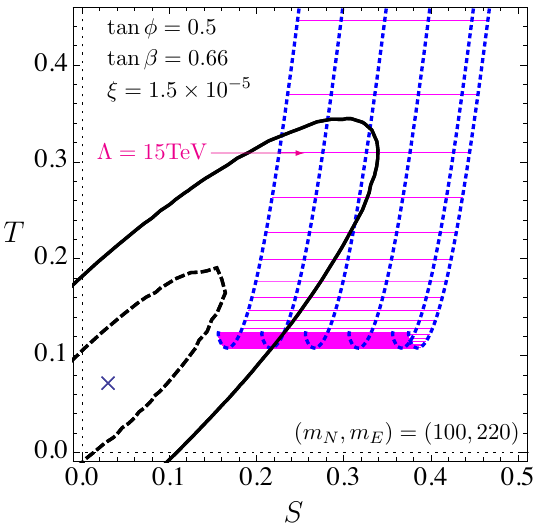} 
%\vspace*{5ex}
\end{minipage}
}
\\
{
\begin{minipage}[t]{0.5\textwidth}
\vspace*{3ex}
\begin{flushleft} (c) \end{flushleft} \vspace*{-5ex}
\includegraphics[scale=0.75]{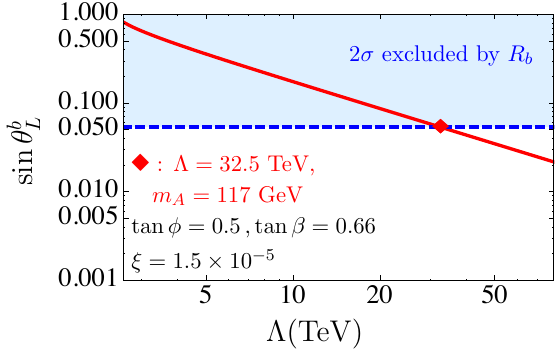} 
%\vspace*{5ex}
\end{minipage}
}
\end{tabular}
\caption[]{ 
(a) ,(b) The Peskin-Takeuchi $S,T$-parameters (a) for $S_{\rm TC} = 0.1$ and (b) for varying $S_{\rm TC}$ and (c) $R_b$ constraints (right panels) for the present model with  $(\tan \phi,\tan \beta) = (0.5,0.66)$, $\xi = 1.5 \times 10^{-5}$ and $(m_N,m_E)=(100 \,\GeV , 220 \,\GeV)$. The allowed region for $m_A$ is $117 \,\GeV \leq m_A \leq 262 \,\GeV$ for $S_{\rm TC} = 0.1$. 
\label{ST-present-model-light}}
\end{center}
\end{figure}

In the model spectrum
%, since the Techniquarks do not carry QCD charge, 
there are two objects which can be discovered in the same decay mode as the SM higgs boson: the lightest CP-odd neutral PNGB $A^0$ or the lightest CP-even neutral boson $h^0$, which are composed of 3rd and 4th family quarks.
% which is different from the analysis in \cite{Chivukula:2011ue,Holdom:2012pw}. Thus we concentrate on the $A^0$.
To study this possibility, we set parameters as $(\tan \phi, \tan \beta) = (0.5,0.66)$ with $\xi = 1.5 \times 10^{-5}$. 
We take this choice of parameters simply as an example to illustrate the viability of the model in light of recent LHC data, and
leave a more thorough scan of the parameter space for future work. With this choice of parameters we can solve the gap equations (\ref{gapeq-U-af}) and (\ref{gapeq-D-af}) only for $2.6 \,\TeV \leq \Lambda \leq 80 \,\TeV$. We show the $S,T$ parameters constraint and  $R_b$ constraint for these cases in Fig.\ref{ST-present-model-light}. The $99.73 \% \cl$ allowed region by the Peskin-Takeuchi $S,T$-parameters (left panels in Fig.\ref{ST-present-model-light}) is $14.6 \TeV \leq \Lambda \leq 80 \TeV$ which corresponds to $56 \,\GeV \leq m_A \leq 262 \,\GeV$ for $(m_N, m_E)=(100 \,\GeV , 220 \,\GeV)$. The allowed region by the $R_b$ constraints is $32.5 \,\TeV \leq \Lambda$ which corresponds to $117 \,\GeV \leq m_A$. Thus, in the present model with these parameters, the allowed region for $m_A$ is $117 \,\GeV \leq m_A\leq 262 \,\GeV$. In this range, mass of the lightest CP-even higgs: $h^0$ is $336 \,\GeV \leq m_h\leq 390 \,\GeV$. 

Next, we will consider how this spectrum of light $A^0$ and $h^0$ is constrained by the LHC results for the SM higgs.
On one hand, the SM-higgs boson whose mass with in this range quoted above, decays mainly to $WW/ZZ$. Hence, we should compare the $h^0$ with the results for the SM-higgs search in $h \to WW \to l\nu jj$ at the ATLAS \cite{ATLAS:2011ae}, $h \to WW \to l\nu l\nu$ channel at the CMS \cite{CMSWW-13Dec2011} and $h \to ZZ \to 4l$ channel at the ATLAS \cite{Collaboration:2012sm} and the CMS \cite{CMSZZ-13Dec2011}.  On the other hand, similarly to the ordinary two higgs doublet model, the $A^0$ here does not have any coupling with $WW/ZZ$ at the tree level, so it is natural to concentrate only on the $A^0 \to \gamma \gamma$ channel. Thus with the above parameter choices, we consider the ratios defined as 
\beq
R_{gg \to \varphi \to X}
\equiv
\kappa^\varphi_{gg} \times
\frac{{\rm Br}(\varphi \to X)}{{\rm Br}(h_{\rm SM} \to X)}\,,
\eeq 
where $\varphi = h^0,A^0$, ${\rm Br}(\varphi \to X) \equiv \Gamma(\varphi \to X)/\Gamma_{\rm tot}$, 
\beq
\kappa^\varphi_{gg}
=
\frac{\sigma(gg \to \varphi)}{\sigma(gg \to h_{\rm SM}) }
=
\frac{\Gamma(\varphi \to gg)}{\Gamma(h_{\rm SM} \to gg) }\,,
\eeq
In the SM, $\Gamma(h_{\rm SM} \to gg)$ is given by \cite{Gunion:1989we}
\beq
\Gamma(h_{\rm SM} \to gg)
=
\frac{\alpha^2_s m^3_h}{32 \pi^3 v^2_{\rm EW}} 
\left| 
\sum_{i=t,b} \tau_i A^{(h)}(\tau_i)
\right|^2\,,
\eeq
and we will use the values of ${\rm Br}(h_{\rm SM} \to X)$ given in \cite{LHCHiggsCrossSectionWorkingGroup:2012vm}. 
The total decay widths, $\Gamma_{\rm tot}(\varphi)$ for $m_{\varphi} < 2 m_{t',b'} (\simeq {\rm few}\, \TeV)$ are given by 
\beq
\Gamma_{\rm tot}(h^0)
\!\!&=&\!\!
\Gamma(h^0 \to gg) + \Gamma(h^0 \to \gamma\gamma)
+\Gamma(h^0 \to t\bar{t}) + \Gamma(h^0 \to b\bar{b}) \nonumber\\[1ex]
&&
+\Gamma(h^0 \to WW) + \Gamma(h^0 \to ZZ)
+\Gamma(h^0 \to A^0A^0) + \Gamma(h^0 \to A^0 Z)  
\,,
\eeq
%%%%%%%%%%%
\beq
\Gamma_{\rm tot}(A^0)
\!\!&=&\!\!
\Gamma(A^0 \to gg) + \Gamma(A^0 \to \gamma\gamma)
+\Gamma(A^0 \to t\bar{t}) + \Gamma(A^0 \to b\bar{b})
\,.
\eeq
The decay widths appearing in above equations are explicitly given in Appendix \ref{decaywidths}.

 In Fig.\ref{h-mass-br}, with model parameters $(\tan \phi, \tan \beta) = (0.5,0.66)$ with $\xi = 1.5 \times 10^{-5}$ and within the region allowed by Fig.\ref{ST-present-model-light}, we show in panel (a) the results for the dynamical higgs masses with respect to $m_A (\GeV)$ in the range $117 \,\GeV \leq m_A \leq \, 262 \GeV$, and in panel (b) the branching ratio ${\rm Br}(h \to X)$ with respect to $m_h (\GeV)$ in the range $336 \,\GeV \leq m_h \leq 390 \,\GeV$.  As one can see from Fig.\ref{h-mass-br}(a), in the present case, $h^0$ can decay mainly into $A^0A^0$ 
%as the same as the top-higgs in \cite{Chivukula:2011ag} 
but $h^0$ can not decay into $H^+H^-/H^\pm W^\mp$. 
%which is different from \cite{Chivukula:2011ag}. 
Moreover, as one can read off from Fig.\ref{h-mass-br}(b), if the $h^0 \to A^0A^0$ channel is kinematically allowed, this decay channel is the dominant decay mode, i.e. $h^0 \to WW/ZZ$ channel is suppressed. This is different from the SM-higgs case where $h^0_{\rm SM} \to WW/ZZ$ channel is dominant in this mass range of $h^0$.%
\begin{figure}[h]
\begin{center}
\begin{tabular}{cc}
{
\begin{minipage}[t]{0.5\textwidth}
\begin{flushleft} (a) \end{flushleft} \vspace*{-5ex}
\includegraphics[scale=0.6]{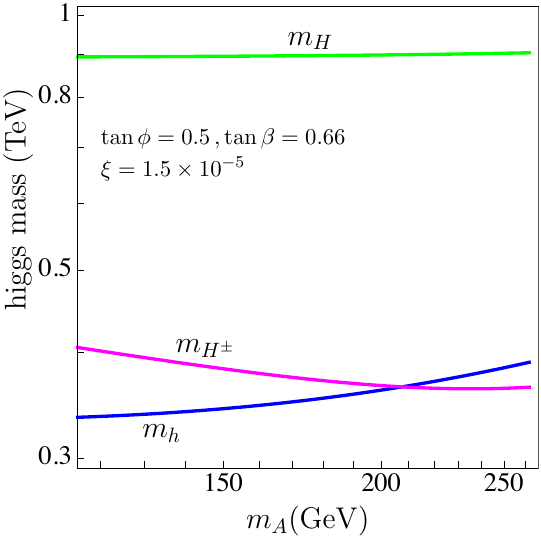} 
%\vspace*{3ex}
\end{minipage}
}
{
\begin{minipage}[t]{0.5\textwidth}
\begin{flushleft} (b) \end{flushleft} \vspace*{1ex}
\includegraphics[scale=0.75]{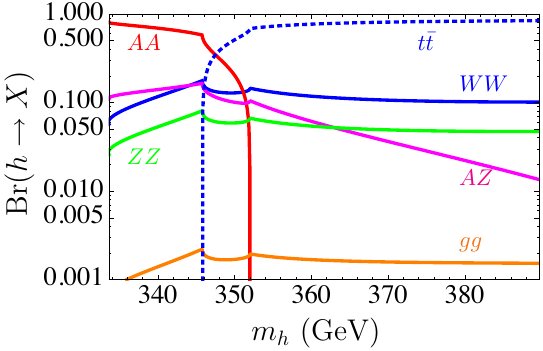} 
%\vspace*{5ex}
\end{minipage}
}
\end{tabular}
\caption[]{ 
(a) the dynamical results for higgs masses and (b) the branching ratio ${\rm Br}(h \to X)$ in the present model with $(\tan \phi, \tan \beta) = (0.5,0.66)$ with $\xi = 1.5 \times 10^{-5}$ in the range of the allowed region by Fig.\ref{ST-present-model-light}. (a): the dynamical results  and the horizontal axis is the $m_A (\GeV)$ with $117 \,\GeV \leq m_A \leq \, 262 \GeV$. (b): the branching ratio of $h^0$ with $336 \,\GeV \leq m_h \leq 390 \,\GeV$. 
\label{h-mass-br}}
\end{center}
\end{figure}%

\begin{figure}[h]
\begin{center}
\begin{tabular}{cc}
{
\begin{minipage}[t]{0.5\textwidth}
%\begin{flushleft} (c) \end{flushleft} \vspace*{-5ex}
\includegraphics[scale=0.75]{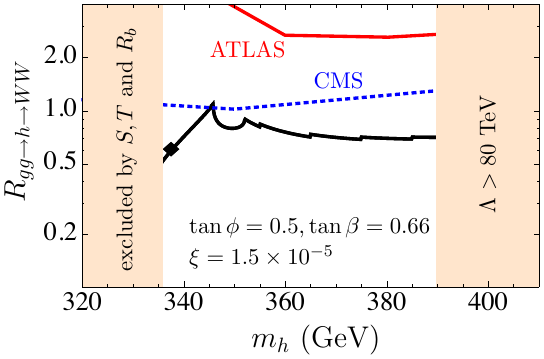} 
%\vspace*{3ex}
\end{minipage}
}
{
\begin{minipage}[t]{0.5\textwidth}
%\begin{flushleft} (d) \end{flushleft} \vspace*{-5ex}
\includegraphics[scale=0.75]{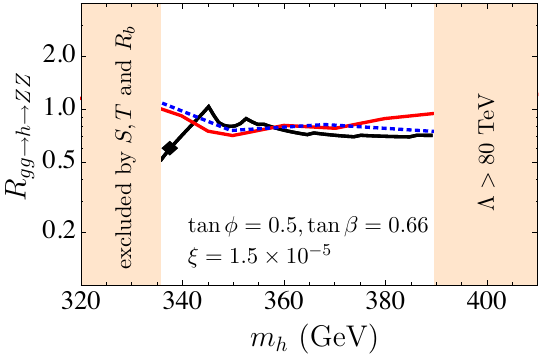} 
%\vspace*{5ex}
\end{minipage}
}
\end{tabular}
\caption[]{ 
$R_{gg \to h \to WW}$(left) and $R_{gg \to h \to ZZ}$(right) in the present model with $(\tan \phi, \tan \beta) = (0.5,0.66)$ and $\xi = 1.5 \times 10^{-5}$. The lower/black solid lines in each figure correspond to $R_{gg \to h \to WW/ZZ}$ in the present model, the upper/red solid lines correspond to $95 \% \cl$ observed upper limit on $R_{gg \to h \to WW/ZZ}$ for the SM higgs boson decay into $WW/ZZ$ at the ATLAS \cite{ATLAS:2011ae,Collaboration:2012sm} and  the blue/dotted lines correspond to $95 \% \cl$ observed upper limit on $R_{gg \to h \to WW/ZZ}$ for the SM higgs boson decay into $WW/ZZ$ at the CMS \cite{CMSWW-13Dec2011,CMSZZ-13Dec2011}. In both figures $\blackdiamond$ implies the point at $\Lambda = 35.3\,\TeV$ corresponding to $(m_A,m_h) = (126\,\GeV, 338\,\GeV)$. $\Lambda > 80 \,\TeV$ implies that the gap equations (\ref{gapeq-U-af}) and (\ref{gapeq-D-af}) do not have any solutions. 
\label{Constraint-LHC-hVV}}
\end{center}
\end{figure}

Based on above results, we show $R_{gg \to h \to WW/ZZ}$ with the allowed range of $m_h$ in Fig.\ref{Constraint-LHC-hVV}. In the present case, gluon fusion process is enhanced compared with the SM higgs case, $\kappa^h_{gg} \simeq 4$. However, the ${\rm Br}(h^0 \to WW/ZZ)$ is suppressed as ${\rm Br}(h^0 \to WW/ZZ)/{\rm Br}(h^0_{\rm SM} \to WW/ZZ) \simeq 0.18$ for $336 \,\GeV \leq m_h \leq 390 \,\GeV$. Thus, the present model gives $R_{gg \to h \to WW/ZZ} \simeq 0.8$ which should be compared with the LHC results  \cite{ATLAS:2011ae,Collaboration:2012sm,CMSWW-13Dec2011,CMSZZ-13Dec2011}. Fig.\ref{Constraint-LHC-hVV} shows that light CP-even higgs $h^0$ with $342 \,\GeV \leq m_h \leq 357 \,\GeV$ is  excluded at $95 \% \cl$ which corresponds to $41.5 ,\TeV \leq \Lambda \leq 55.5\,\TeV$ and $146 \,\GeV \leq m_A \leq 190 \,\GeV$ in the present case.

\begin{figure}[h]
\begin{center}
%\begin{tabular}{cc}
%{
%\begin{minipage}[t]{0.5\textwidth}
%\begin{flushleft} (a) \end{flushleft} \vspace*{-5ex}
%\includegraphics[scale=0.6]{ATLAS-constraint-tss.pdf} 
%\vspace*{2ex}
%\end{minipage}
%}
%{
%\begin{minipage}[t]{0.5\textwidth}
%\begin{flushleft} (b) \end{flushleft} \vspace*{-5ex}
%\includegraphics[scale=0.6]{CMS-constraint-tss.pdf} 
%\vspace*{2ex}
%\end{minipage}
%}
%\\
%{
%\begin{minipage}[t]{0.5\textwidth}
%\begin{flushleft} (c) \end{flushleft} \vspace*{-5ex}
\includegraphics[scale=0.75]{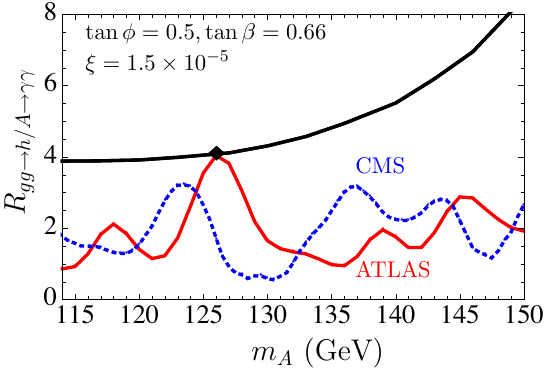} 
%\vspace*{3ex}
%\end{minipage}
%}
%minipage}
%}
%\end{tabular}
\caption[]{ 
$R_{gg \to h/A \to \gamma\gamma}$ in the present model with $(\tan \phi, \tan \beta) = (0.5,0.66)$ with $\xi = 1.5 \times 10^{-5}$. The upper/black solid line corresponds to $R_{gg \to A \to \gamma\gamma}$ in the present model, the lower/red solid line corresponds to $95 \% \cl$ observed upper limit on $R_{gg \to h \to \gamma\gamma}$ in the SM higgs boson decay into two photons at the ATLAS \cite{ATLAS-13Dec2011} and the  dotted/blue line corresponds to $95 \% \cl$ observed upper limit on $R_{gg \to h \to \gamma\gamma}$ at the CMS  \cite{CMS-13Dec2011}. $\blackdiamond$ implies the point at $\Lambda = 35.3\,\TeV$ corresponding to $(m_A,m_h) = (126\,\GeV, 338\,\GeV)$.
\label{Constraint-LHC-h2p}}
\end{center}
\end{figure}

Thus we find the CP-odd higgs $A^0$ in this model with $117 \GeV \leq m_A \leq 146\,\GeV$ is allowed in light of the present data. Based on this result, we consider $R_{gg \to A \to \gamma\gamma}$ in the present model with $(\tan \phi, \tan \beta) = (0.5,0.66)$ and $\xi = 1.5 \times 10^{-5}$.  
In Fig.\ref{Constraint-LHC-h2p}, we show $R_{gg \to A \to \gamma\gamma}$ in the present case together with the $95 \% \cl$ observed upper limit on $R_{gg \to h_{\rm SM} \to \gamma\gamma}$ at the ATLAS \cite{ATLAS-13Dec2011} and the CMS \cite{CMS-13Dec2011}. This means that the excess around $126 \,\GeV$ of only $\gamma\gamma$-channel in the ATLAS and CMS data implies a signal of neutral top-pion i.e. CP-odd higgs $A^0$ in the present model since the neutral top-pion is EW-gaugephobic. 
%
%It is easy to read from Fig.\ref{Constraint-LHC-h2p} that $A_0$ with $m_A \leq 150 \GeV$ in the present model is invisible in the ATLAS data and $A^0$ with $128 \GeV \leq m_A \leq 132 \GeV\,,145 \GeV \leq m_A \leq 148 \GeV$ is excluded by the CMS data in $\gamma\gamma$-channel but $A^0$ with $m_A \geq 150 \GeV$ may be discovered in this channel at the LHC. Of course, this result may be changed if we consider the mixing with the NGBs coming from the MWT sector in which the lightest pion $\pi^0_{\rm H4G}$ is a mixture of $A^0$ and $\pi^0_{\rm TC}$. %so $\pi^0_{\rm H4G}$ is allowed to decay into $bb/cc/\tau^+\tau^-$ for $m_{\pi_{\rm H4G}} \leq 2m_t$  via tree level interaction which are not allowed in the case of $A^0$. 
In addition, in a generic walking TC theory it is possible that the low energy spectrum contains a techni-dilaton \cite{Yamawaki:1985zg}, and this may mainly decay into $\gamma\gamma$ \cite{Matsuzaki:2012gd}. Further analysis can be carried out by taking into consideration contributions from both light PNGBs from the Technicolor sector as well  as the techni-dilaton in the context of our model. 

\section{Conclusions}

We have considered a model where electroweak symmetry breaking is due to top seesaw and technicolor dynamics, and which also contains novel matter fields with quantum numbers of SM matter fields.  Hence, as an interesting aside, this model serves as an example of a model with a new type of hybrid fourth generation: while the leptons of the fourth generation transform as usual chiral fermions under the weak gauge group, the QCD quarks of the fourth generation transform as vectorlike fermions. We have carried out a full phenomenology analysis of this model and established the spectrum as well as the constraints from the oblique corrections and $Zb\bar{b}$-vertex. We also discussed implications of this model in light of present and future LHC data. In particular we have showed that the excess in $\gamma\gamma$-channel recently observed at the LHC could be explained as due to decay of a light CP-odd Higgs boson of this model. 

Our model can be viewed as a low energy effective theory for the composite states arising from the strong dynamics underlying the top seesaw and technicolor, together with a hybrid fourth generation matter. As an explicit microscopic realization we have considered a consistent extension of Minimal Walking Technicolor (MWT).

Several novel features have been uncovered by our analysis. As in the original MWT model, the fourth generation leptons play an essential role. First, they provide a clear phenomenological signatures, and, second, they enter the precision constraints and are important for the viability of the model. In the present model, which extend the MWT model to account also for the masses of heavy quarks, the SM like fourth generation lepton doublet is accompanied by vectorlike fourth generation of QCD quarks. Hence, this model is very different from the usual extensions of the SM with a sequential fourth generation. We determined the plausible mass spectra for this {\em hybrid} fourth generation.

The scalar sector is richer than in MWT, since here, in addition to the Technicolor compositeness, there are five physical composite higgs particles which arise due to the dynamical condensates between third and fourth generation QCD quarks. We determined the mass spectrum of these new scalars. In addition there will be additional scalars from the MWT degrees of freedom and also states of higher spin. In this paper we have taken these to be heavy and decoupled from the low energy spectrum, and only considered the effect of technicolor through its contribution to the electroweak symmetry breaking condensate.

In light of early data from LHC, strong dynamics remains as a viable explanation as the mechanism of the electroweak symmetry breaking to be uncovered in further measurements at the LHC. Any technicolor model requires an extension 
towards the generation of masses for the (B)SM matter fields. Hence, these type of models are likely to receive further attention in the future.

%We analysed in detail the constraints from the precision measurements, and found the model viable. Here we emphasize the role played by the fourth generation leptons required to exist due to internal consistency of the underlying MWT model.

%In the future we will investigate the collider signatures of this model for the phenomenologically viable portions of the parameter space. 

%%%%%%%%%%%%%%%%%%%%%%%%%%%%%%%%%%%%%%%%%%
%%%%%%%%%%%%%%%%%%%%%%%%%%%%%%%%%%%%%%%%%%
%%%%%%%%%%%%%%%%%%%%%%%%%%%%%%%%%%%%%%%%%%
%
%
%
\appendix
\def\thesection{\Alph{section}}
\renewcommand{\theequation}{\Alph{section}.\arabic{equation}}
\setcounter{equation}{0}
%
%
%%%%%%%%%%%%%%%%%%%%%%%%%%%%%%%%%%%%%%%%%%
%%%%%%%%%%%%%%%%%%%%%%%%%%%%%%%%%%%%%%%%%%

\section{The EW gauge interaction of quarks}

In the present model, the fourth generation leptons are chiral representation under the electroweak gauge group. On the other hand, the fourth generation QCD quarks are in vector representations under the electroweak gauge group. So, the leptons kinetic terms are identical to the ones in the ordinary SM. However, the fourth generation quark gauge interactions are different from the SM, and we have
% and are given by 
%${\cal L}^{\rm 4 th}_{\rm kin} \left.\right|_{\rm quark} \ni {\cal L}_{\rm cc} + {\cal L}_{\rm nc}$ and 
\beq
{\cal L}_{\rm cc} 
\!\!&=&\!\!
g_2 \sum^3_{i=1} \overline{Q^{(i)}_L} \gamma^\mu W_\mu Q^{(i)}_L 
\,,\\[1ex]
{\cal L}_{\rm nc} 
\!\!&=&\!\!
\frac{1}{6} g_1 \sum^3_{i=1} \overline{Q^{(i)}_L} \gamma^\mu  B_\mu Q^{(i)}_L 
+ g_1 \sum^3_{i=1} \overline{Q^{(i)}_R} \gamma^\mu \bpm 2/3 & 0 \\ 0 & -1/3 \epm B_\mu Q^{(i)}_R 
\nonumber\\[1ex]
\!\!&&\!\! 
+ g_1 \overline{Q^{(4)}} \gamma^\mu \bpm 2/3 & 0 \\ 0 & -1/3 \epm B_\mu Q^{(4)}
\,,
\eeq 
where $W_\mu$ $(B_\mu)$ are $SU(2)$ $(U(1))$ gauge bosons fields, $g_{2(1)}$ is short for the $SU(2)$ $(U(1)_Y)$ gauge coupling and $Q^{(i)}_R = \left( U^{(i)}_R \,,\, D^{(i)}_R \right)^T$. 
By using the rotation matrix in Eq. (\ref{quark-mixing}), 
%${\cal L}^{\rm 4 th}_{\rm kin} \left.\right|_{\rm quark}$ 
these interactions can be represented in terms of the mass basis. The resulting vertices are shown in Table.~\ref{vertex-ffEW} 
%
%\beq
%{\cal L}_{\rm cc} 
%= 
%\frac{g_2}{\sqrt{2}} W^+_\mu \sum^4_{\alpha,\beta = 1} {\cal V}_{\alpha \beta} \, \bar{u}^%{(\alpha)}_L \gamma^\mu d^{(\beta)}_L
%+ {\rm h.c.}\,,
%\eeq
%and
%\beq
%{\cal L}_{\rm nc}
%\!\!&=&\!\!
%\frac{g_2}{2} W^3_\mu  \sum^4_{\alpha,\beta=1} {\cal U}^L_{\alpha \beta} \, \bar{u}^{(\alpha)}_L \gamma^\mu u^{(\beta)}_L
%-
%\frac{g_2}{2} W^3_\mu  \sum^4_{\alpha,\beta=1} {\cal D}^L_{\alpha \beta}\, \bar{d}^{(\alpha)}_L \gamma^\mu d^{(\beta)}_L
%\,\nonumber\\[1ex]
%&&
%+
%g_1 B_\mu
%\sum^4_{\alpha,\beta=1} 
%\bar{u}^{(\alpha)} \gamma^\mu 
%\left[
 %\left(-\frac{1}{2} {\cal U}^L_{\alpha \beta}+ \frac{2}{3} \delta_{\alpha \beta} \right) P_L
 %+ \frac{2}{3} \delta_{\alpha \beta} P_R
%\right]
%u^{(\beta)}
%\,\nonumber\\[1ex]
%&&
%+
%g_1 B_\mu
%\sum^4_{\alpha,\beta=1} 
%\bar{d}^{(\alpha)} \gamma^\mu 
%\left[
 %\left(\frac{1}{2} {\cal D}^L_{\alpha \beta} - \frac{1}{3} \delta_{\alpha \beta} \right) P_L
 %-\frac{1}{3} \delta_{\alpha \beta} P_R
%\right]
%d^{(\beta)}\,,
%\eeq
where $P_{L,R} = (1 \mp \gamma_5)/2$ and the matrices ${\cal V},{\cal U},{\cal D}$ are given by
\beq
&&
{\cal V}_{\alpha \beta} \equiv \sum^3_{i=1} U^{(L) *} _{i\alpha} D^{(L)}_{i\beta}
= \delta_{\alpha \beta} - U^{(L) *} _{4\alpha} D^{(L)}_{4\beta}
\,,
\label{def-calV-matrix}
\\[1ex]
&&
{\cal U}^L_{\alpha \beta} \equiv  \sum^3_{i=1} U^{(L) *} _{i\alpha} U^{(L)}_{i\beta}
= \delta_{\alpha \beta} - U^{(L) *} _{4\alpha} U^{(L)}_{4\beta}
\,,
\label{def-calU-matrix}
\\[1ex]
&&
{\cal D}^L_{\alpha \beta} \equiv  \sum^3_{i=1} D^{(L) *} _{i\alpha} D^{(L)}_{i\beta}
= \delta_{\alpha \beta} - D^{(L) *} _{4\alpha} D^{(L)}_{4\beta}
\,,
\label{def-calD-matrix}
\eeq
which are not necessarily unitary.

\begin{table}[h]
\begin{center}
{
\renewcommand\arraystretch{2.5}
\begin{tabular}{| c  || c | c |}
\hline
vertex & $i g_L$ & $i g_R$
\\
\hline
\parbox[c][7.5ex][c]{0ex}{}
$g_2 W^+_\mu \bar{u}^{(\alpha)} d^{(\beta)}$ & $i\,\dfrac{1}{\sqrt{2}} \,{\cal V}_{\alpha \beta} $  
& 0
\\ \hline
\parbox[c][7.5ex][c]{0ex}{}
$g_2 W^3_\mu \bar{u}^{(\alpha)} u^{(\beta)}$ & $i\, \dfrac{1}{2} \, {\cal U}^L_{\alpha \beta} $ 
& 0
\\ \hline
\parbox[c][7.5ex][c]{0ex}{}
$g_2 W^3_\mu \bar{d}^{(\alpha)} d^{(\beta)}$ & $-i\,\dfrac{1}{2} \, {\cal D}^L_{\alpha \beta}$
& 0 
\\ \hline
\parbox[c][10ex][c]{0ex}{}
$g_1 B_\mu \bar{u}^{(\alpha)} u^{(\beta)}$ & 
$i \left[ -\dfrac{1}{2} \, {\cal U}^L_{\alpha \beta}+ \dfrac{2}{3} \, \delta_{\alpha \beta} \right]$
& $i \,\dfrac{2}{3} \delta_{\alpha \beta}$
\\ \hline
\parbox[c][10ex][c]{0ex}{}
$ g_1 B_\mu \bar{d}^{(\alpha)} d^{(\beta)}$ & 
$i \left[\dfrac{1}{2} \, {\cal D}^L_{\alpha \beta} - \dfrac{1}{3} \, \delta_{\alpha \beta} \right]$ 
& $- i\, \dfrac{1}{3} \, \delta_{\alpha \beta}$ 
\\
\hline
\end{tabular}
}
\caption{The quark-electroweak gauge boson couplings in mass eigenbasis of the quarks in the form of $\gamma^\mu \left[ g_L P_L + g_R P_R\right]$ where $P_{L,R} = (1 \mp \gamma_5)/2$. $\alpha,\beta = 1,2,3,4$ and ${\cal V,U,D}$ are given in Eqs.(\ref{def-calV-matrix},\ref{def-calU-matrix},\ref{def-calD-matrix}).} 
\label{vertex-ffEW}
\end{center}
\end{table}%

%%%%%%%%%
%%%%%%%%%
\section{Effective theory parameters}
\setcounter{equation}{0}
\label{matching}
%%%%%%%%%
%%%%%%%%%
The masses, ${\cal M}^2_{\pi}$ and ${\cal M}^2_{\pi\pm}$ in the low energy effective theory are given
in terms of the high energy theory parameters through the computation of Fig.\ref{BHL-bubble} (a) as
\beq
M^2_A = \frac{\xi\Lambda^2}{\cos\beta\sin\beta \sqrt{Z_{\Phi_1}Z_{\Phi_2}}}\,,
\label{dyn-CPodd-higgs}
\eeq
where we have dropped ${\cal O}(\xi^2)$, and 
\beq
%
% charged higgs
%
M^2_{H\pm} \!\!\!&=&\!\!\!
[{\cal M}^2_{\pi\pm}]_{11} \sin^2 \beta 
+ 
[{\cal M}^2_{\pi\pm}]_{22} \cos^2 \beta
+
2 [{\cal M}^2_{\pi\pm}]_{12} \sin\beta\cos\beta 
\label{dyn-charged-higgs}
\,,\\[1ex]
[{\cal M}^2_{\pi\pm}]_{11} \!\!\!&=&\!\!\! 
Z^{-1}_{\Phi_1} 
\left[
\left(1-\frac{3 y^2_{10}}{8 \pi^2} \sum^4_{\alpha,\beta=1} |D^R_{\alpha4}|^2 |U^L_{\beta3}|^2 \right)\Lambda^2 
+ \frac{3 y^2_{10}}{8 \pi^2} \sum^4_{\alpha,\beta=1} |D^R_{\alpha4}|^2 |U^L_{\beta3}|^2 F_1(m_{d\alpha},m_{u\beta})
\right]\,,
\nonumber\\
\label{dyn-ch-11}
\\[1ex]
[{\cal M}^2_{\pi\pm}]_{22} \!\!\!&=&\!\!\! 
Z^{-1}_{\Phi_2} 
\left[
\left(1-\frac{3 y^2_{20}}{8 \pi^2} \sum^4_{\alpha,\beta=1} |D^R_{\alpha3}|^2 |U^L_{\beta4}|^2 \right)\Lambda^2 
+ \frac{3 y^2_{10}}{8 \pi^2} \sum^4_{\alpha,\beta=1} |D^R_{\alpha3}|^2 |U^L_{\beta4}|^2 F_1(m_{d\alpha},m_{u\beta})
\right]\,,
\nonumber\\
\label{dyn-ch-22}
\\[1ex]
[{\cal M}^2_{\pi\pm}]_{12} \!\!\!&=&\!\!\!  
\frac{1}{Z_{\Phi_1} Z_{\Phi_2}} \frac{3 y_{10} y_{20}}{8 \pi^2} 
\sum^4_{\alpha,\beta=1}{\rm Re}[(D^L_{\alpha3} D^{R*}_{\alpha4})(U^L_{\beta3} U^{R*}_{\beta4})] F_0(m_{d\alpha},m_{u\beta})
\label{dyn-ch-12}
\,.
\eeq
Here $F_{0,1}(m,M)$ are given by
\beq
F_0(m,M)
&=& 
m M \left[ \frac{m^2}{m^2 -M^2} \ln \frac{\Lambda^2 +m^2_1}{m^2_1} 
+ \frac{M^2}{M^2 -m^2} \ln \frac{\Lambda^2 +M^2}{M^2} \right]
\,\\[1ex]
&=&
m^2 \left[ \ln \frac{\Lambda^2 +m^2}{m^2} - \frac{\Lambda^2}{\Lambda^2 + m^2} \right]
\quad , \quad
(\text{for $m = M$})\,,
\\[2ex]
F_1(m^2,M^2)
&=&
\frac{m^4}{m^2 -M^2} \ln \frac{\Lambda^2 +m^2}{m^2} 
+ \frac{M^4}{M^2 -m^2} \ln \frac{\Lambda^2 +M^2}{M^2}
\,\\[1ex]
&=&
m^2\left[ 2\ln \frac{\Lambda^2 +m^2}{m^2} - \frac{\Lambda^2}{\Lambda^2 + m^2}\right] 
\quad , \quad
(\text{for $m = M$})\,.
\eeq

Similarly, for the quartic couplings  $\lambda'_{t,b}$ and $\lambda_{1,2}$  we obtain through the computation of Fig.\ref{BHL-bubble} (b) the relations
\beq
\lambda'_b = \xi \cdot 2\lambda_1 \sqrt{\frac{Z_{\Phi_1}}{Z_{\Phi_2}}}
\quad,\quad
\lambda'_t = \xi \cdot 2\lambda_2 \sqrt{\frac{Z_{\Phi_2}}{Z_{\Phi_1}}}\,,
\label{relation-prime12}
\eeq
and 
\beq
%
% CP-even higgs
%
2 \lambda_1 v^2_1
\!\!\!&=&\!\!\!
Z^{-1}_{\Phi_1} 
\left[
\begin{aligned}
&
\left(1-\frac{3 y^2_{10}}{8 \pi^2} \right)\Lambda^2 \\
&
+ \frac{3 y^2_{10}}{8 \pi^2} \sum^4_{\alpha,\beta=1}
\left\{ {\rm Re}[(D^L_{\alpha3} D^{R*}_{\alpha4})(D^L_{\beta3} D^{R*}_{\beta4})] F_0(m_{d\alpha},m_{d\beta})
+
 |D^L_{\alpha3}|^2|D^R_{\beta4}|^2 F_1(m_{d\alpha},m_{d\beta})
\right\}
\end{aligned}
\right]\,,
\nonumber\\
\label{dyn-CPeven-1}
\\[1ex]
2 \lambda_2 v^2_2 &=& 
Z^{-1}_{\Phi_2} 
\left[
\begin{aligned}
&\left(1-\frac{3 y^2_{20}}{8 \pi^2} \right)\Lambda^2  \\
&
+ \frac{3 y^2_{20}}{8 \pi^2}\sum^4_{\alpha,\beta=1}  
\left\{{\rm Re}[(U^L_{\alpha3} U^{R*}_{\alpha4})(U^L_{\beta3} U^{R*}_{\beta4})] F_0(m_{u\alpha},m_{u\beta})
+ |U^L_{\alpha3}|^2|U^R_{\beta4}|^2 F_1(m_{u\alpha},m_{u\beta}) \right\}
\end{aligned}
\right]\,.
\label{dyn-CPeven-2}
\nonumber\\[1ex]
\eeq

%%%%%%%%%%%%%%%%%%%%%%%%%%%%%%%%%%%%%%%%%%
%%%%%%%%%%%%%%%%%%%%%%%%%%%%%%%%%%%%%%%%%%
\section{Formulae for integrals}
\setcounter{equation}{0}
\label{integral}
%%%%%%%%%%%%%%%%%%%%%%%%%%%%%%%%%%%%%%%%%%
%%%%%%%%%%%%%%%%%%%%%%%%%%%%%%%%%%%%%%%%%%

We define the following integrals 
\beq
{\cal F}_0(q^2,m^2,M^2) 
\!\!\!&=&\!\!\!
\int^1_0 \!\!\!dz 
\ln R(q^2,m^2,M^2)
\,,\\[1ex]
{\cal F}_1(q^2,m^2,M^2) 
\!\!\!&=&\!\!\!
\int^1_0 \!\!\!dz 
\left[R(q^2,m^2,M^2) +  z(z-1)q^2 \right] \ln R(q^2,m^2,M^2) 
\,, \\[1ex]
{\cal F}_2(q^2,m^2,M^2) 
\!\!\!&=&\!\!\!
\int^1_0 \!\!\!dz \,
R(q^2,m^2,M^2) \ln R(q^2,m^2,M^2)
\,,\\[2ex]
R(q^2,m^2,M^2)
\!\!\!&=&\!\!\!
 z(z-1)q^2 + (1-z) m^2 + zM^2.
\eeq
These are evaluated as
\beq
{\cal F}_0(q^2,m^2,M^2) 
\!\!\!&=&\!\!\!
 \frac{1}{2}\ln (xy) - 2 -\frac{f(x,x)+f(y,y)}{4}
+ \frac{1}{2}\left[ \chi_- (x,y) - \frac{1}{2} \theta_-(x,y)\right]
\,,\\[1ex]
{\cal F}_0(0,m^2,M^2) 
\!\!\!&=&\!\!\!
 \frac{1}{2}\ln (xy)
-\frac{1}{2} \theta_-(x,y)
\,,\\[1ex]
{\cal F}_1(q^2,m^2,M^2) 
\!\!\!&=&\!\!\! 
\frac{5}{9} - \frac{x+y}{3}-\frac{1}{6}\ln(xy) + \frac{x\ln x + y \ln y }{2}
-\frac{x-1}{12}f(x,x) - \frac{x-1}{12}f(y,y) 
\nonumber\\[1ex]
&& \!\!\!
+ \frac{1}{2}\left[ \chi_+ (x,y) - \frac{1}{2} \theta_+(x,y)\right]
\,,\\[1ex]
{\cal F}_2(q^2,m^2,M^2) 
\!\!\!&=&\!\!\! 
 \frac{x\ln x + y \ln y }{2}- \frac{1}{4} \theta_+(x,y) 
+ {\cal F}'_1(x,y) 
+ \frac{(x-y)^2 - (x+y)}{2}{\cal F}'_0(x,y)
\,,\nonumber\\ \\[1ex]
{\cal F}_1(0,m^2,M^2) 
\!\!\!&=&\!\!\! 
{\cal F}_2(0,m^2,M^2) 
=
 \frac{x\ln x + y \ln y }{2} - \frac{1}{4} \theta_+(x,y) 
\,,\\[1ex]
{\cal F}'_0(q^2,m^2,M^2)
\!\!\!&=&\!\!\!
{\cal F}_0 (q^2,m^2,M^2) - {\cal F}_0 (0,m^2,M^2) 
\,,\\[1ex]
{\cal F}'_1(q^2,m^2,M^2)
\!\!\!&=&\!\!\!
{\cal F}_1 (q^2,m^2,M^2) - {\cal F}_1 (0,m^2,M^2) 
\,,
\eeq
where $\theta_\pm(x,y)$ and $\chi_\pm(x,y)$ are given by \cite{Lavoura:1992np}
\beq
\theta_+(x,y) 
\!\!\!&=&\!\!\! 
x + y - \frac{2 x y}{x - y} \ln\frac{x}{y}\,,
\\[2ex]
\theta_-(x,y)
\!\!\!&=&\!\!\!
\frac{x+y}{x-y}\ln\frac{x}{y} -2
\,,\\[2ex]
\chi_+(x,y) 
\!\!\!&=&\!\!\! 
\frac{x+y}{2} -\frac{(x-y)^2}{3} 
+ \left[ \frac{(x-y)^3}{6} - \frac{x^2+y^2}{2(x-y)}\right] \ln\frac{x}{y}
\\[1ex]
&&+\frac{x-1}{6}f(x,x) + \frac{x-1}{6}f(y,y)+ \left[ \frac{1}{3} - \frac{x+y}{6} - \frac{(x-y)^2}{6} \right] f(x,y)\,,\nonumber
\\[1ex]
\chi_-(x,y)
\!\!\!&=&\!\!\!
2 + \left[ x-y-\frac{x+y}{x-y} \right] \ln \frac{x}{y} + \frac{f(x,x) + f(y,y)}{2} - f(x,y) 
\,,\\[2ex]
\theta_\pm(x,x) \!\!\!&=&\!\!\! \chi_\pm(x,x) = 0
\,,\\[2ex]
f(x,y)
\!\!\!&=&\!\!\!
\left\{ \begin{array}{ll}
- 2 \sqrt{\Delta} \left[ \arctan \dfrac{x-y+1}{\sqrt{\Delta}} - \arctan \dfrac{x-y-1}{\sqrt{\Delta}}\right] & \text{for $\Delta > 0$} \\[1ex]
0 & \text{for $\Delta = 0$} \\[1ex]
\sqrt{-\Delta} \ln \dfrac{x+y-1+\sqrt{-\Delta}}{x+y-1-\sqrt{-\Delta}}  & \text{for $\Delta < 0$} \\[1ex]
\end{array} \right.,
\\[1ex]
\Delta \!\!\!&=&\!\!\! -1 + 2(x+y) -(x-y)^2\,, 
\eeq
in which $\theta_\pm(x,x),\chi_\pm(x,x) = 0$.
In this appendix, $x,y$ are given by
\beq
x= \begin{cases}
\dfrac{m^2}{q^2} & \text{for\,\,} q^2 \neq 0 \\[3ex]
\dfrac{m^2}{M^2_Z} & \text{for\,\,} q^2 \neq 0 
\end{cases}
\quad , \quad
y= \begin{cases}
\dfrac{M^2}{q^2} & \text{for\,\,} q^2 \neq 0 \\[3ex]
\dfrac{M^2}{M^2_Z} & \text{for\,\,} q^2 \neq 0 
\end{cases}\,.
\eeq

%%%%%
%%%%%
\section{Decay widths of the Higgs}
\setcounter{equation}{0}
\label{decaywidths}
%%%%%
%%%%%

The decay widths for $\varphi$ into two gauge bosons are given by
%%%% two gluons 
\beq
\Gamma(\varphi \to gg)
\!\!&=&\!\!
\frac{\alpha^2_s m^3_{\varphi}}{32 \pi^3 v^2_{\rm EW} } 
\left| 
\sum_{i=t,t'} R^{(u)}_i (\varphi) \tau_i A^{(\varphi)}(\tau_i)
+\sum_{i=b,b'} R^{(d)}_i (\varphi)  \tau_i A^{(\varphi)}(\tau_i)
\right|^2
\,,\label{decay-Agg}
\eeq
%%%% two photons
\beq
\Gamma(\varphi \to \gamma\gamma)
\!\!&=&\!\!
\frac{\alpha^2_e m^3_\varphi}{64 \pi^3 v^2_{\rm EW}} 
\left| 
3 \cdot \left( \frac{2}{3}\right)^2 \sum_{i=t,t'} R^{(u)}_i(\varphi) \tau_i A^{(\varphi)}(\tau_i)
+ 3 \cdot \left( -\frac{1}{3}\right)^2  \sum_{i=b,b'} R^{(d)}_i(\varphi)  \tau_i A^{(\varphi)}(\tau_i)
\right|^2\,, \nonumber \\
&&
\label{decay-A2p}
\eeq
%%%% two W
\beq
\Gamma(h^0 \to WW) 
=
\sin^2(\beta - \alpha) \times \frac{m^2_h}{16 \pi v^2_{\rm EW}} \sqrt{1- x_W} \left[1-x_W + \frac{3}{4}x^2_W \right]\,,
\label{decay-h2W}
\eeq
%%%% two Z
\beq
\Gamma(h^0 \to ZZ) 
=
\sin^2(\beta - \alpha) \times \frac{m^2_h}{16 \pi v^2_{\rm EW}} \sqrt{1- x_Z} \left[1-x_Z + \frac{3}{4}x^2_Z \right]\,,
\label{decay-h2Z}
\eeq
where $x_V \equiv 4m^2_h/M^2_V\,,(V=W,Z)$.% and they are the same as the representation in the 2HDM \cite{Gunion:1989we}, 
The decay widths for $\varphi$ into two fermions are given by
%%%%
\beq
\Gamma(\varphi\to t\bar{t})
\!\!&=&\!\!
\left[R^{(u)}_3 (\varphi)\right]^2 \times 
\frac{3 m_\varphi m^2_t }{8 \pi v^2_{\rm EW}} 
\left[ 1 - \frac{4m^2_t}{m^2_\varphi} \right]^{1/2}
\,,\label{decay-Att}
\eeq
%%%%
\beq
\Gamma(\varphi \to b\bar{b})
\!\!&=&\!\!
 \left[R^{(d)}_3 (\varphi)\right]^2 \times 
\frac{3 m_\varphi m^2_b}{8 \pi v^2_{\rm EW}}
\left[ 1 - \frac{4m^2_b}{m^2_\varphi} \right]^{1/2}\,,
\label{decay-Abb}
\eeq
where $R^{(u,d)}_i(\varphi)$ is given by 
\beq
R^{(u)}_i(\varphi) 
\!\!\!&=&\!\!\! 
\left\{ \begin{array}{ll}
\dfrac{\Sigma_U}{m_i} \left[ U^{L*}_{i3} U^R_{i4} \right] \dfrac{1}{\cos \phi} \dfrac{\cos\alpha}{\sin\beta}\,,
& \,\, \text{for $\varphi = h$}\,, \\[2ex]
\dfrac{\Sigma_U}{m_i} \left[ U^{L*}_{i3} U^R_{i4} \right] \dfrac{\cot \beta}{\cos \phi}\,,
& \,\, \text{for $\varphi = A$}\,.
\end{array} \right.
\\[2ex]
R^{(d)}_i(\varphi) 
\!\!\!&=&\!\!\! 
\left\{ \begin{array}{ll}
\dfrac{\Sigma_D}{m_i} \left[ D^{L*}_{i3} D^R_{i4} \right] \dfrac{1}{\cos \phi}\dfrac{\sin\alpha}{\cos\beta}\,,
& \,\, \text{for $\varphi = h$}\,, \\[2ex]
\dfrac{\Sigma_D}{m_i} \left[ D^{L*}_{i3} D^R_{i4} \right] \dfrac{\tan \beta}{\cos \phi} 
& \,\, \text{for $\varphi = A$}\,.
\end{array} \right.
\eeq
and $A^{(\varphi)}(\tau_i)\,,(\tau_i \equiv 4m^2_i/m^2_\varphi)$ is given by 
\beq
A^{(\varphi)}(\tau_i) 
\!\!\!&=&\!\!\! 
\left\{ \begin{array}{ll}
1 + (1-\tau_i)A(\tau_i) & \,\, \text{for $\varphi = h$}\,, \\[2ex]
 A(\tau_i) & \,\, \text{for $\varphi = A$}\,,
\end{array} \right.
\eeq
where
\beq
A(\tau_i)
\!\!\!&=&\!\!\!
\left\{ \begin{array}{ll}
[\arcsin(1/\sqrt{\tau_i})]^2 & \text{for $\tau_i > 1$}\,, \\[2ex]
-\dfrac{1}{4} \left[ \log \dfrac{1+\sqrt{1-\tau_i}}{1-\sqrt{1-\tau_i}} - i \pi \right]^2  & \text{for $\tau_i \leq 1$} \,.\\[1ex]
\end{array} \right.
\eeq
The above decay widths Eqs.(\ref{decay-Agg},\ref{decay-A2p},\ref{decay-Att},\ref{decay-Abb}) are similar as in the two Higgs doublet model \cite{Gunion:1989we}, but taking into account also the fourth family quarks contributions.
% under $R^{(u)}_i(h) \to \cos \alpha/\sin \beta$, $R^{(d)}_i(h) \to \sin \alpha/\cos \beta$, $R^{(u)}_i(A) \to \cot \beta$ and $R^{(d)}_i(A) \to \tan \beta$. 
%This is because, as one can read off from Eq.(\ref{reno-yukawa-H4G}), the yukawa couplings in the present model are presented by a form as $y = \Sigma/(\sqrt{2} v)$ where $\Sigma$ is dynamical mass and varies together with the cutoff $\Lambda$. 
In addition, if kinematically allowed, $h^0$ decays into $A^0A^0$ and $A^0Z^0$ and these decay widths are given by
%%%% h -> AA
\beq
\Gamma(h^0\to A^0A^0)
=
\frac{\lambda^2_{hAA}}{32 \pi m_h} 
\sqrt{1-\frac{4 m^2_A}{m^2_h}}\,,
\label{decay-hAA}
\eeq
%%%% h -> AZ
\beq
\Gamma(h^0\to A^0Z)
=
\cos^2(\beta - \alpha) \times \frac{m^3_h}{16 \pi v^2_{\rm EW}} 
\left[ 1-\frac{(m_A - M_Z)^2}{m^2_h}\right]^{3/2} 
\left[ 1-\frac{(m_A + M_Z)^2}{m^2_h}\right]^{3/2}\,,
\label{decay-hAZ}
\eeq
where $\lambda_{hAA}$ is represented as
\beq
\lambda_{hAA}
\!\!\!&=&\!\!\!
\lambda_b v_1 \sin \alpha \sin^2_\beta
- \lambda_t v_2 \cos \alpha \cos^2 \beta 
- \frac{1}{2} \lambda_{tb} \sin(\beta - \alpha) \left[ v_1 \cos \beta + v_2 \sin \beta\right]
\nonumber\\[1ex]
&& 
- \frac{1}{2}\lambda'_b \left[ \left( v_1 \cos \beta - v_2 \sin \beta \right) \sin \alpha - v_1 \sin(\beta - \alpha) \sin \beta \right]
\nonumber\\[1ex]
&& 
- \frac{1}{2}\lambda'_t \left[ \left( v_1 \cos \beta - v_2 \sin \beta \right) \cos \alpha - v_2 \sin(\beta + \alpha) \cos \beta \right]\,,
\eeq
and follows from the potential Eq.(\ref{higgs-potential}) in the mass eigenbasis for each higgs boson.

%%%%%%%%%%

\end{document}